\documentclass[reprint,onecolumn]{revtex4-2}
\usepackage{multirow,booktabs,hyperref,graphicx,babel,blindtext,caption,subcaption,tikz,amsmath}
\usepackage[inline]{enumitem}
\newcommand{\figref}[1]{Fig. (\ref{#1})}
\newcommand{\eqnref}[1]{Eqn\eqref{#1}}
\newcommand{\fex}{Fe$_{x}$Rh$_{100-x}$}
\newcommand{\feo}{Fe$_{50}$Rh$_{50}$}
\newcommand{\fem}{Fe$_{49}$Rh$_{51}$}
\newcommand{\fep}{Fe$_{51}$Rh$_{49}$}
\newcommand{\jij}{$\mathcal{J}_{ij}$}
\begin{document}

\title{Tuning electronic and magnetic properties of FeRh alloy by chemical and physical method}
\author{Greeshma R}
\author{Rudra Banerjee}
\email{bnrj.rudra@gmail.com}
\affiliation{
	Department of Physics and Nanotechnology, SRM Institute of Science and Technology, Kattankulathur, Tamil Nadu, 603203, India
}
\begin{abstract}
	The electronic, magnetic, and thermodynamic properties of ordered and chemically disordered FeRh alloy is studied using
	\textit{ab-initio} methods. The equiatomic \feo~composition is reported for both ordered and disordered phases.  Chemically
	disordered \fex~is reported and the effect of disorder on electronic and magnetic properties is discussed. Further, We have
	reported the effects of stress and strain in both the order and disorder phases. The result is only for the cubic phase and no
	distortion has been taken into consideration. This study is motivated by the recent resurgence in FeRh study motivated by the
	fact that the barocaloric properties can be possible to sustain over the cycle. Hence, we have discussed the properties of
	\fex~with chemical disorder and pressure together, to gain an insight into the compound effect and the interplay between them.
\end{abstract}
\maketitle

\section{Introduction} 
\label{sec:introduction}
The century old Joule-Thomson effect for cooling technology is not only inefficient but also emits environmentally hazardous
by-products.  In the last decade caloric materials(CM) has been staged as a feasible alternative due to its environment friendly cooling
technology and improved efficiency\cite{Swartzendruber_1984}. The caloric materials show significant adiabatic temperature change
($\Delta T$) or isothermal entropy change ($\Delta S$) near phase boundary under the applied magnetic, electric or mechanical field,
corresponding to magnetocaloric(MC)\cite{Staunton_2014}, electrocaloric(EC) or mechanocaloric(mC) elasto(eC) or barocaloric(BC))
effects\cite{Moya_2014}. Though the magnetocaloric effect is believed to be observed first in 19th century\cite{Smith_2013}, the
field really took off with the observation of `giant' mC\cite{Rodriguez_1980}, MC\cite{Nikitin_1990} and
EC\cite{Mischenko_2006} in 1980s and subsequent decades.

Various CMs, particularly based on Lanthanides(specially, Gd)\cite{Fujita_2003} and transition metal (specially,
Mn\cite{Dung_2011} and Fe) materials have shown impressive caloric properties. The multicaloric materials, where materials can
show caloric properties stimulated by one or multiple fields as described above, are rare and far between. While there are
theoretical understanding of multicaloric systems\cite{Liu_2014}, experiments are mostly constrained by individual
fields\cite{murthy_2014}. Off-stoichiometric FeRh system is one of the known systems  showing inverse-MC(iMC), BC and EC. FeRh
shows giant iMC near room temperature\cite{Nikitin_1990} at $350K$ from a first-order antiferromagnetic (AFM) to ferromagnetic (FM)
transition. The transition results around 1\% volume expansion, leading to the BC property of
FeRh\cite{Stern_Taulats_2014,stern_2015,Moruzzi_1992,Antoni_2014}. FeRh exhibits around $\Delta S \approx 20JK^{-1}kg^{-1}$ and
$\Delta T\approx 20K$ at 2T, making it an excellent CM. But, FeRh also shows broad hysteresis\cite{franco_2012,stern_2015},
leading to a degradation of caloric properties over refrigeration cycles. However, Liu \textit{et.al} has shown\cite{Liu_2016}
that it is possible to enhance the property over the cycles. Qiao \textit{et.al}\cite{Qiao_2022} has recently shown the effect of
growth on MC properties of FeRh.

This renewed interest in off-stoichiometric FeRh system demands a detailed look into the electronic and magnetic structure of
\fex, where $x$ is the percentage of concentration at a site ranging from 49-51, with varying pressure.
In this article, we have delineated our systematic study of \fex, subject to hydrostatic pressure.  We have
studied the variation in the electronic structure, magnetic exchange and magnetic transition temperature of FeRh alloy. We
carried out the study of this alloy in both FM and AF domain. The study unveiled that the ground state \fex~system is
ferromagnetic in nature(\figref{fig:magopt}). Hence, in this report, we have focused the complete study on FM behaviour of \fex.

\section{Methods} 
\label{sec:methods}
The off-stoichiometric composition and disordered alloy system is best managed using multiple scattering Green's function and
Coherent Potential Approximation (CPA) respectively. The electronic and magnetic calculations in this study are performed using
SPRKKR code\cite{sprkkr,Ebert_2011}. The Perdew, Burke and Ernzerhoff(PBE) generalized gradient approximation (PBE-GGA) is used
for exchange-correlational functional\cite{Perdew_1996}. We performed first Brillouin zone integrations on the 2500 grid of
$k$-points and the energy convergence criteria had been set to $10^{-5}Ry$. The full potential spin polarised scalar relativistic
calculations with angular momentum cut-off $l_{max}=2$ has been performed.

\begin{figure}[!ht]
	\centering
	\begin{subfigure}[b]{.3\columnwidth}
		\includegraphics[width=.9\linewidth]{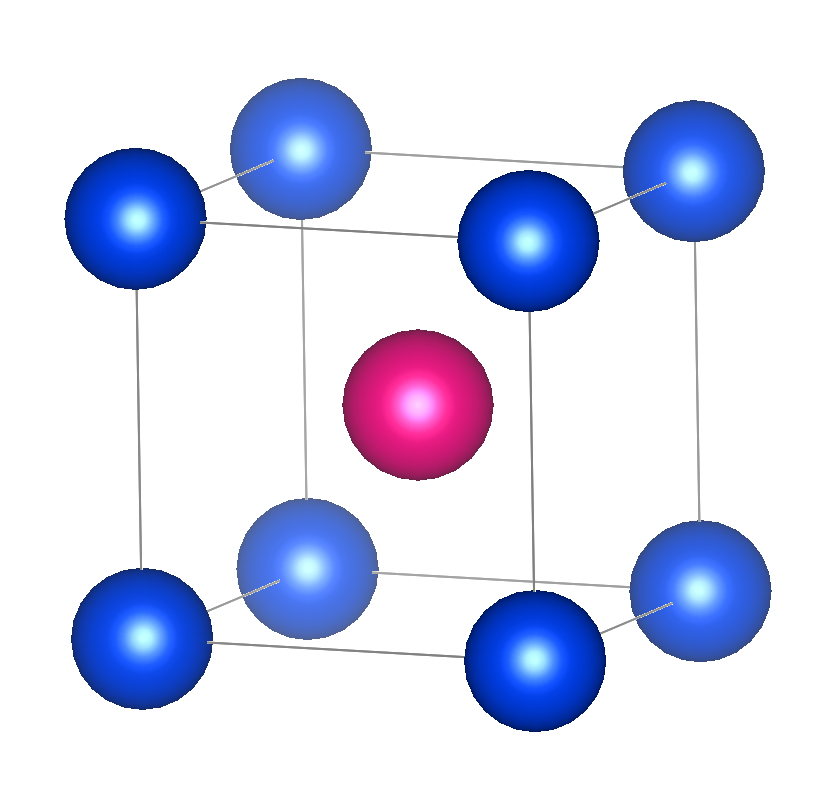}
		\caption{FeRh unit cell with CsCl structure.}
		\label{fig:str}
	\end{subfigure}
	\begin{subfigure}[b]{.3\columnwidth}
		\includegraphics[width=.9\linewidth]{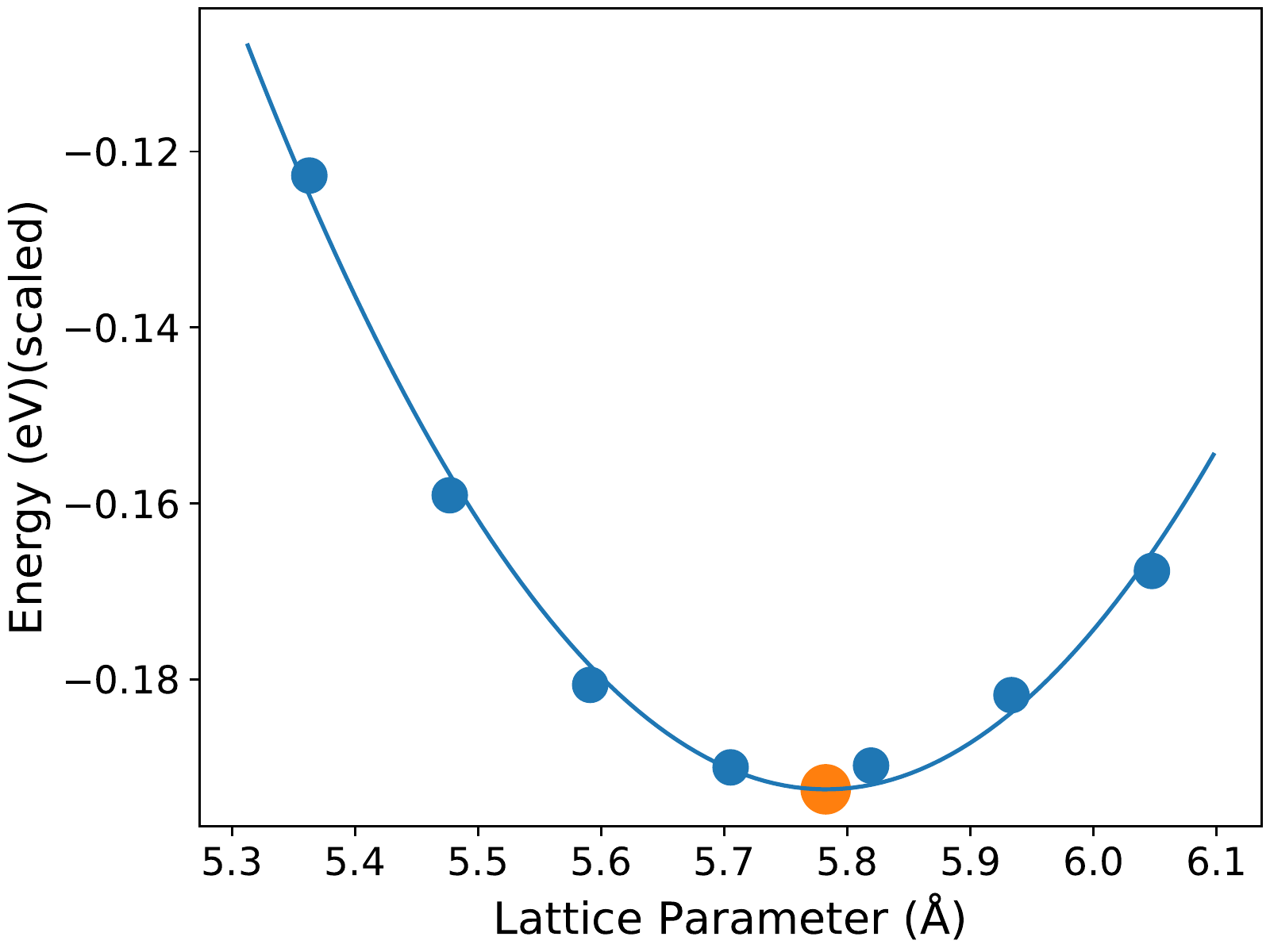}
		\caption{Calculation of minimum energy Lattice parameter for \feo.}
		\label{fig:opt}
	\end{subfigure}
	\begin{subfigure}[b]{.3\columnwidth}
		\includegraphics[width=.9\linewidth]{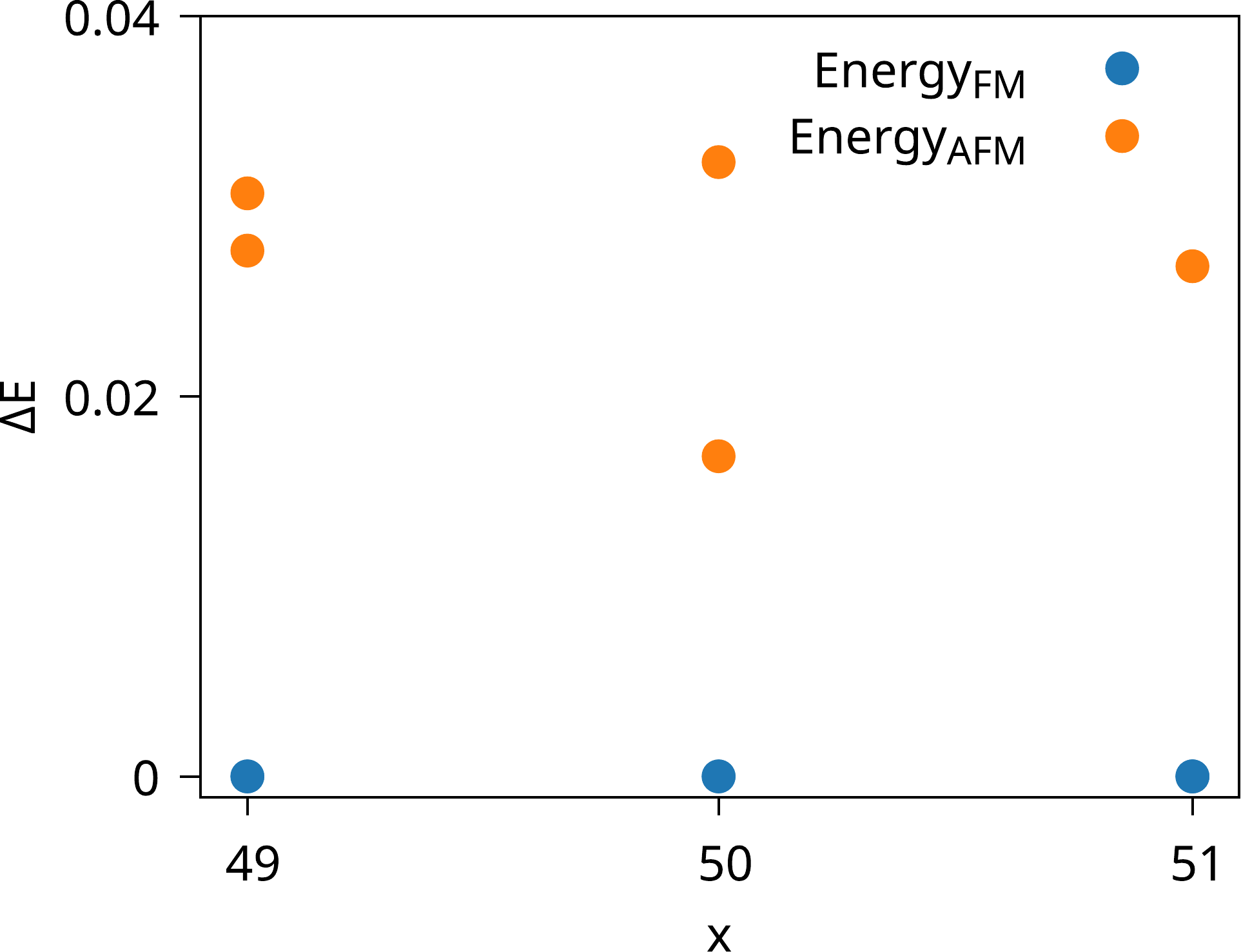}
		\caption{Energy difference of \fex in FM and AFM magnetic structure.}
		\label{fig:magopt}
	\end{subfigure}
	\caption{Initial information of \fex. (\subref{fig:str}) shows the \fex~structure. In ordered structure Fe and Rh occupies
	either corner (0,0,0) or central (.5,.5,.5) site. In disordered systems both (0,0,0) and (.5,.5,.5) sites are occupied by Fe
	and Rh as per concentration. (\subref{fig:opt}) shows the energy minimisation of lattice structure to find the ground state
	lattice parameter. (\subref{fig:magopt}) shows the energy difference between AFM and FM state. In the $y$-axis, $\Delta
		E=E_{mag}-E_{FM}$ is plotted, where $E_{mag}$ is either $E_{FM}$ or $E_{AFM}$.}
\end{figure}

Initially we determined the lattice parameter with minimum energy for the \feo~structure using the following
procedure; i) Procured the lattice parameter from materials project database. ii) Computed the scf of the system, by varying
lattice parameter ranging from 94\% to 106\%, similar calculation is done for each lattice parameters. iii) The curve of lattice
parameter vs energy is fitted by quadratic polynomial equation. iv) The minimum of the curve is the optimised lattice parameter.
For off-stoichiometric structures, we took the optimised lattice parameter from the stoichiometric calculations as the starting
point and followed the steps mentioned above. The optimisation curve of ordered \feo~is shown in the \figref{fig:opt}.

The magnetic exchange energy ($\mathcal{J}_{ij}$) was evaluated to understand the effects of magnetic interactions using the
Lichtenstein formula\cite{Liechtenstein_1984,Terasawa_2019}.
\begin{equation}
	\begin{aligned}
		\mathcal{J}_{ij} & = \frac{1}{4\pi}\int d\varepsilon f(\beta(\varepsilon-\varepsilon_F)) Im Tr[ \hat{\Delta}_i
		\hat{T}^{ij}_{\uparrow} \hat{\Delta}_j \hat{T}^{ji}_{\downarrow}]                                              \\
		                 & = \frac{1}{4\pi}\int d\varepsilon f(\beta(\varepsilon-\varepsilon_F))Im Tr[
				\hat{G}^{+}_{\uparrow}(\varepsilon) \hat{P}_i\hat{G}^{+}_{\downarrow}(\varepsilon) \hat{P}_j]
	\end{aligned}
	\label{eq:lich}
\end{equation}
where, $\hat{T}^{ij}_{\sigma}$ is the scattering path operator for sites $i$ and $j$, $ \hat{\Delta}\equiv \hat{t}_{i\uparrow}-
	\hat{t}_{j\downarrow}$ is the single site scattering matrix $ \hat{t}_{i\sigma}$ at site $i$, $f(x)=1/(e^x+1)$, $\beta=1/k_BT$
and $\varepsilon_F$ is the Fermi energy, $\hat{G}^{+}_{\sigma}(\varepsilon)$ is the retarded Green's function of the spin
$\sigma$ in unperturbed state and $\hat{P}_i \equiv \hat{H}_{i\uparrow} - \hat{H}_{i\downarrow} $ where, $\hat{H}_{i\sigma}$ is
the on-site term of the Hamiltonian for spin $\sigma$ at site $i$.  The $\mathcal{J}_{ij}$ is reckoned by the energy difference
due to the infinitesimal change in magnetic direction, as mapped out from  \eqnref{eq:lich}.
Finally, the Curie temperature is computed using Mean Field Theory(MFT):
\begin{equation}
	k_B T_C = \frac{2}{3} \mathcal{J}_{ij}
	\label{eq:jij}
\end{equation}
where, $\mathcal{J}_{ij}$ is the largest eigen value of the determinant\cite{Liechtenstein_1987,Witte_2016}.

\section{Result} 
\label{sec:result}
FeRh has CsCl structure in ordered phase(B2-phase), with space group 221 (Pm$\bar{3}$m), as shown in \figref{fig:str}. In the
ordered phase, Fe and Rh occupies $(0,0,0)$ and $\left(\frac{1}{2},\frac{1}{2},\frac{1}{2}\right)$ sublattice, respectively. In
chemically disordered phase, \feo~stabilises is bcc-A2 structure\cite{Witte_2016}. In this phase, both Fe and Rh is occupying both
sublattices with equal probability. \fep~and \fem~systems are also disordered with (0,0,0) site is Fe rich or deficient,
respectively, as described in \figref{fig:str}.
\subsection{Ordered FeRh system}
\label{sec:ordered}
\subsubsection{Electronic Structure}
\label{ssub:ord_electronic_structure}
\begin{figure}[ht!]
	\begin{subfigure}[b]{.3\columnwidth}
		\includegraphics[width=\columnwidth]{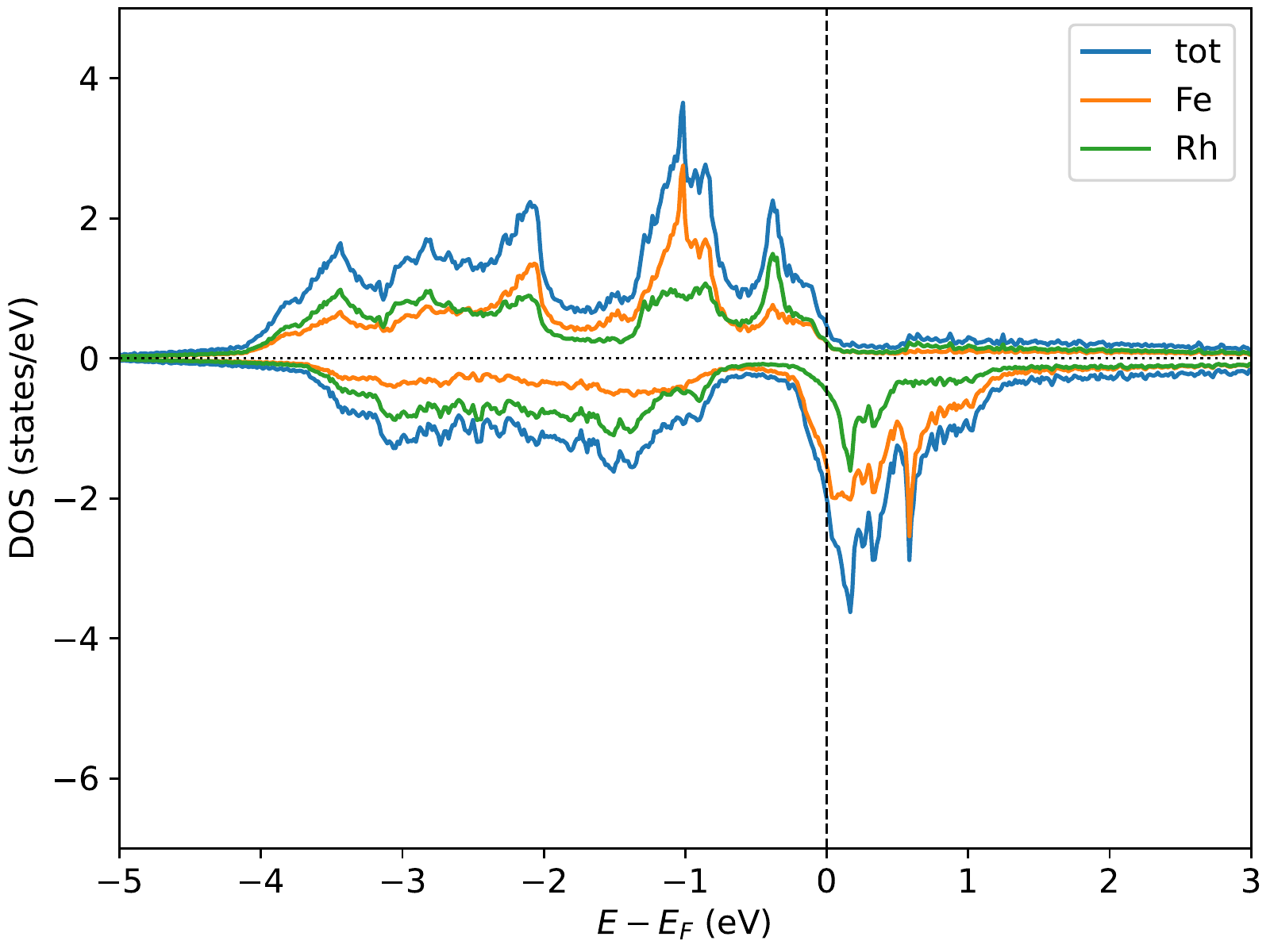}
		\caption{DOS for 6\% strained lattice parameter($a$).}
		\label{fig:dos_ord_94}
	\end{subfigure}
	\begin{subfigure}[b]{.3\columnwidth}
		\includegraphics[width=\columnwidth]{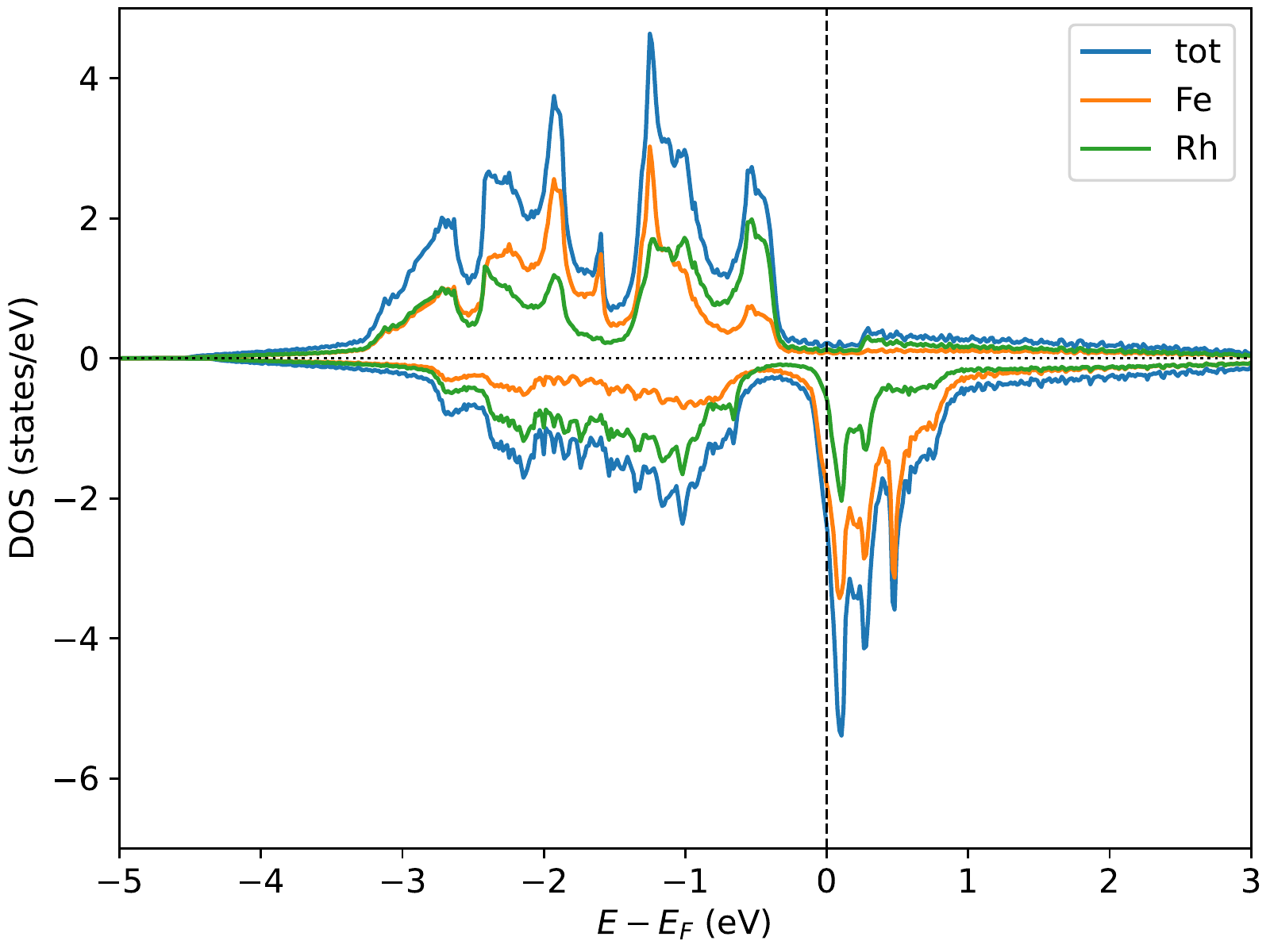}
		\caption{DOS for optimised lattice parameter($a$).}
		\label{fig:dos_ord_opt}
	\end{subfigure}
	\begin{subfigure}[b]{.3\columnwidth}
		\includegraphics[width=\columnwidth]{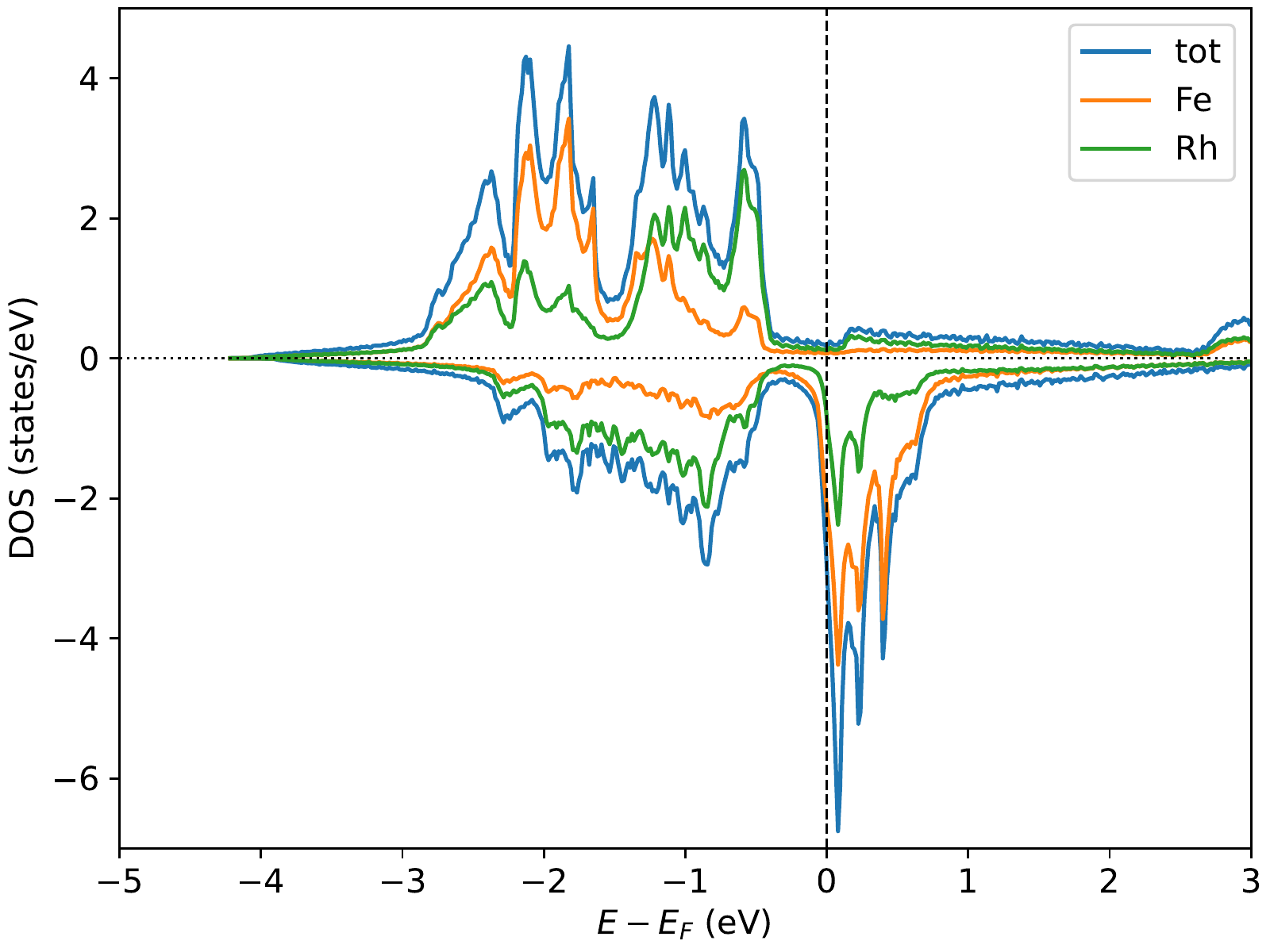}
		\caption{DOS for 6\% stretched lattice parameter($a$).}
		\label{fig:dos_ord_106}
	\end{subfigure}
	\caption{The electronic structure of ordered \feo.}
	\label{fig:ord_e}
\end{figure}

The total and atoms-projected density of states (DOS) for this system is shown in the \figref{fig:ord_e}.
\figref{fig:dos_ord_opt} shows the DOS of \feo~at optimised lattice parameter.
\figref{fig:dos_ord_94} is DOS of \feo~
with 6\% compressed lattice parameter, resembling very high hydrostatic pressure.
Likewise, \figref{fig:dos_ord_106} shows the electronic structure of \feo~ with 6\% stretch.

The \figref{fig:ord_e} shows the DOS of \feo~with (\subref{fig:dos_ord_94})strained, (\subref{fig:dos_ord_opt})optimised and
(\subref{fig:dos_ord_106})stretched lattice parameter. The pseudogap around 1eV down in minority spin channel in optimised
lattice structure is originated from the $3d$ hybridisation of Fe and Rh. The peak just below the $E_F$ in
\figref{fig:dos_ord_94} in the majority spin channel, originating from the hybridisation of the same orbitals shows the
Jahn-Teller instability\cite{Roy_2016,Bordel_2012}. With the increase in lattice parameter $a$, the peak moves further down from
$E_F$, stabilising against the possible Jahn-Teller distortion. The volume-preserving stretch/strain with varying $c/a$ ratio
may show the enhancement of Jahn-Teller distortion, but the possibility within cubic phase decreases with increase in $a$.

In real systems, such a high hydrostatic pressure or stretch is not only nonphysical, but may have other unforeseen
effects as well.  Here we have used this to check the extreme conditions. In all the following discussions, this 6\%
stretch/strain has been considered without further explanation.
\subsubsection{Magnetic Properties}
\label{ssub:ord_magnetic_interactions}
Element resolved magnetic moments $\mu_{Fe}, \mu_{Rh}$ and $\mu_{FeRh}$  of ordered \feo~is shown in Table (\ref{tab:magp}). For
optimised and strained structure, $\mu_{Fe}>3\mu_B$ as already noted by previous study\cite{Kudrnovsk__2015}.
\begin{figure}[ht!]
	\begin{subfigure}[b]{.3\columnwidth}
		\includegraphics[width=\columnwidth]{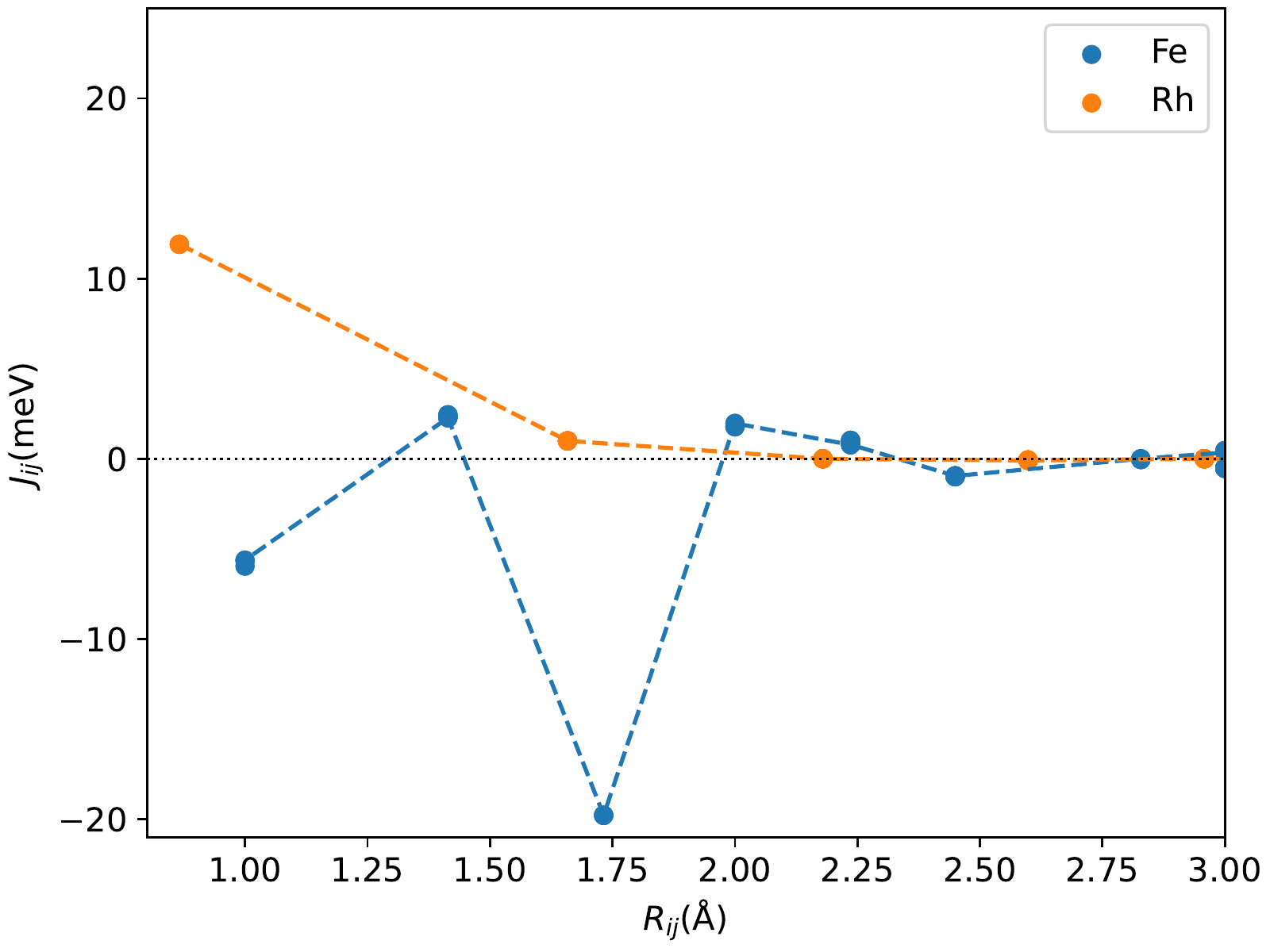}
		\caption{$\mathcal{J}_{ij}$ for 6\% strained lattice parameter($a$).}
		\label{fig:jij_ord_94}
	\end{subfigure}
	\begin{subfigure}[b]{.3\columnwidth}
		\includegraphics[width=\columnwidth]{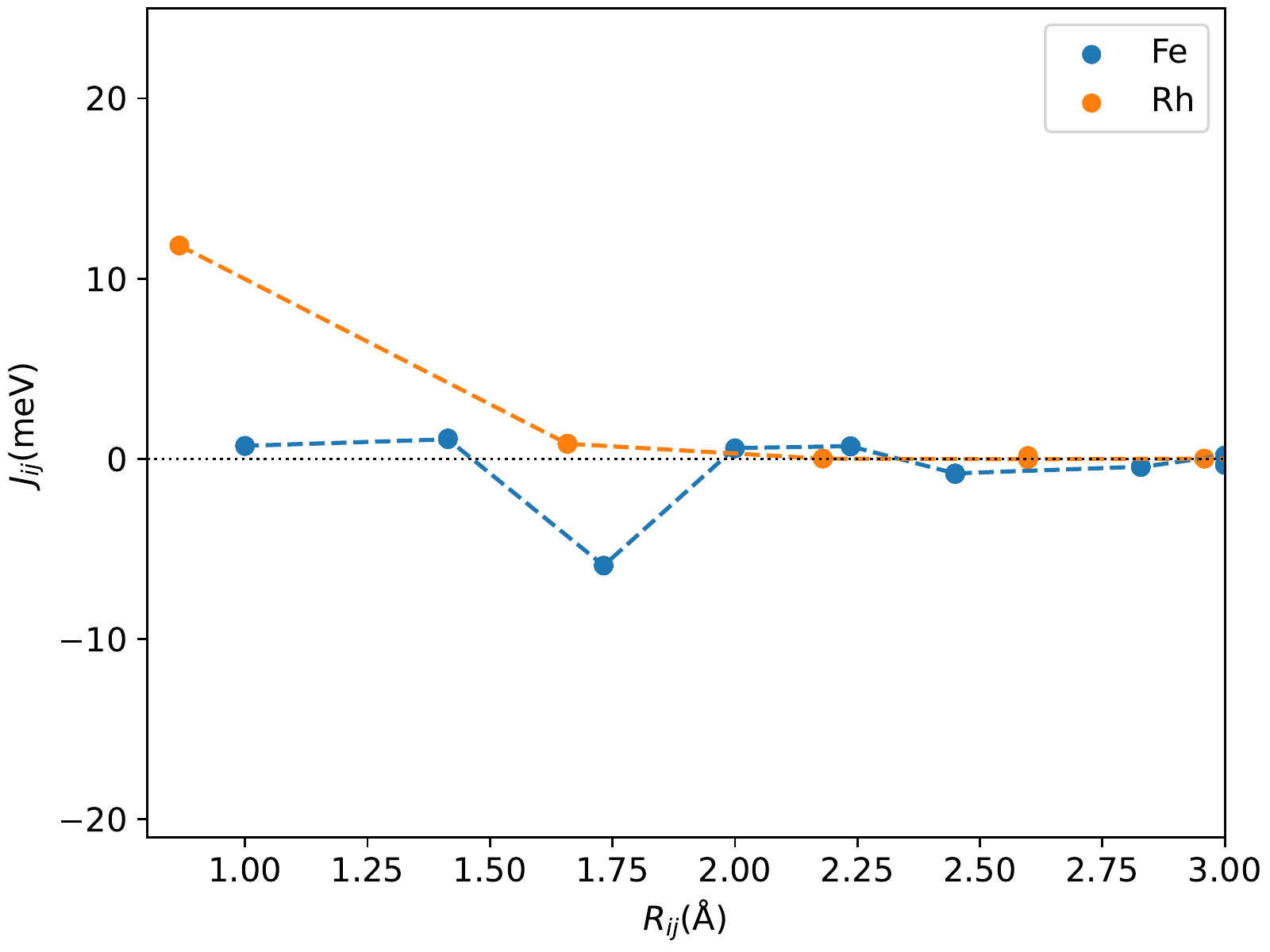}
		\caption{$\mathcal{J}_{ij}$ for optimised lattice parameter($a$).}
		\label{fig:jij_ord_opt}
	\end{subfigure}
	\begin{subfigure}[b]{.3\columnwidth}
		\includegraphics[width=\columnwidth]{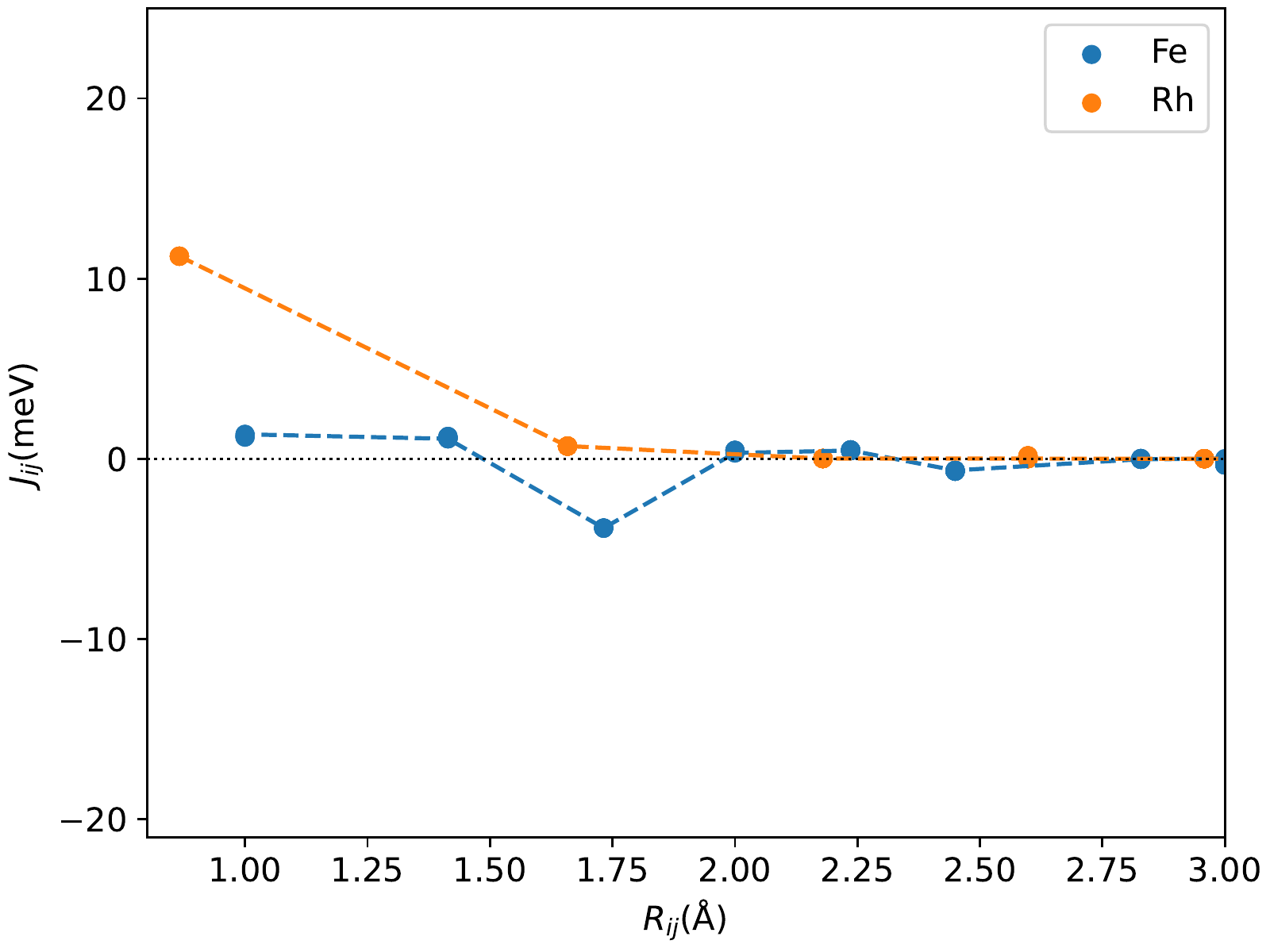}
		\caption{$\mathcal{J}_{ij}$ for 6\% stretched lattice parameter($a$).}
		\label{fig:jij_ord_106}
	\end{subfigure}
	\caption{Magnetic interaction ($\mathcal{J}_{ij}$) of ordered \feo.}
	\label{fig:ord_m}
\end{figure}
\figref{fig:ord_m} shows the magnetic interactions of ordered \feo~with Fe of (0,0,0) sublattice at the center. The stretched configuration in \figref{fig:jij_ord_94} is interesting as it is showing very strong alternating positive and
negative interaction, leading to a spin-glass like configuration. The glassy phase in FeRh is already known for metastable fcc
phase\cite{Navarro_1996,Chen_2017,Barua_2013}, but to the best of our knowledge, we are the first to observe spin-glass effect in
these 221 systems using pressure only.  In optimised or stretched system, the glassy behaviour is not observed, and dominant
interactions are ferromagnetic.


\subsection{Disordered FeRh system}
\label{sec:disorder}
\begin{figure}[ht!]
	\centering
	\begin{subfigure}{\columnwidth}
		\begin{subfigure}[b]{.3\columnwidth}
			\renewcommand\thesubfigure{\alph{subfigure}.1}
			\includegraphics[width=\columnwidth]{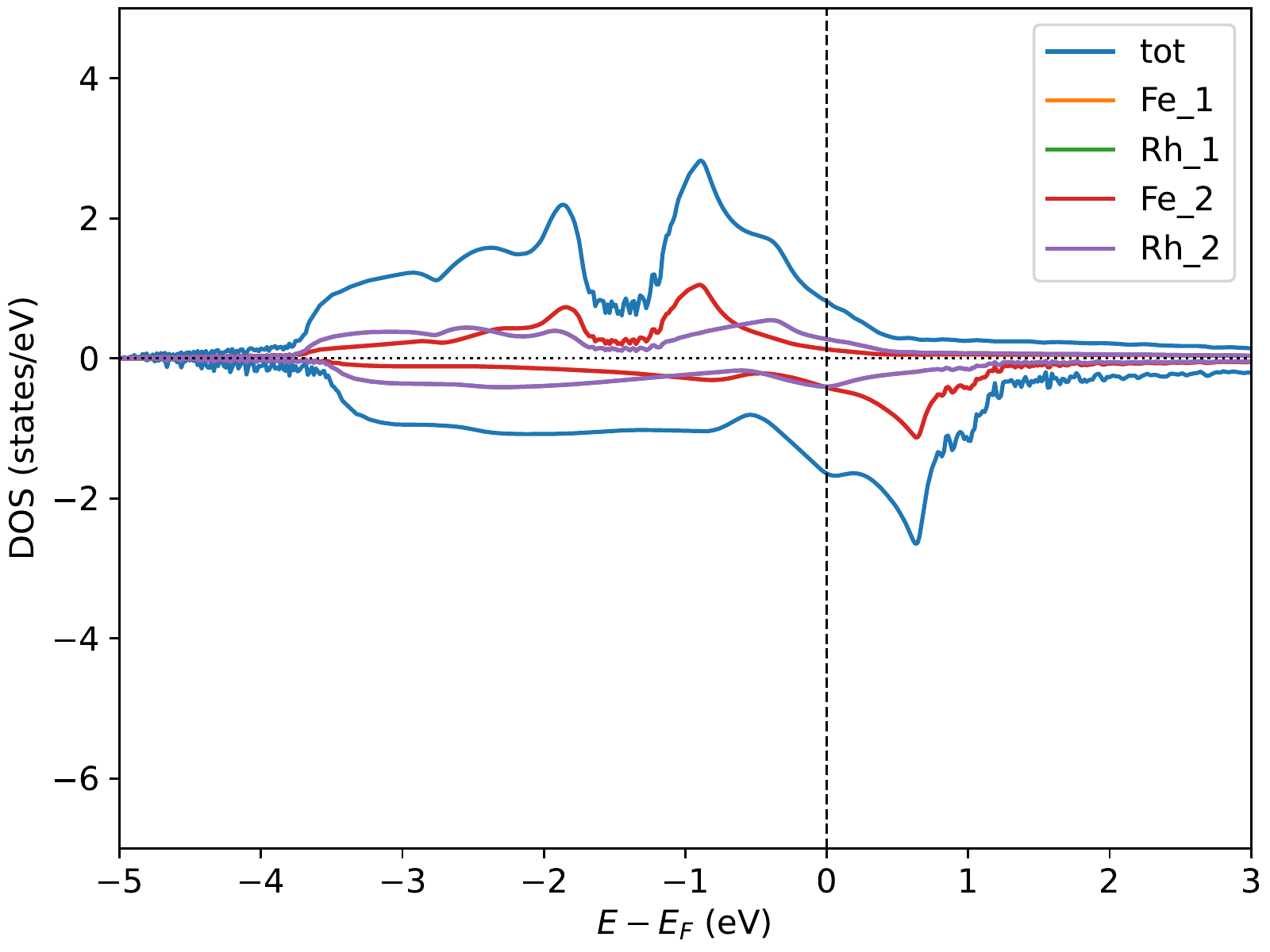}
			\caption{DOS for 6\% strained lattice parameter($a$).}
			\label{fig:dos_d49_94}
		\end{subfigure}
		\begin{subfigure}[b]{.3\columnwidth}
			\addtocounter{subfigure}{-1}
			\renewcommand\thesubfigure{\alph{subfigure}.2}
			\includegraphics[width=\columnwidth]{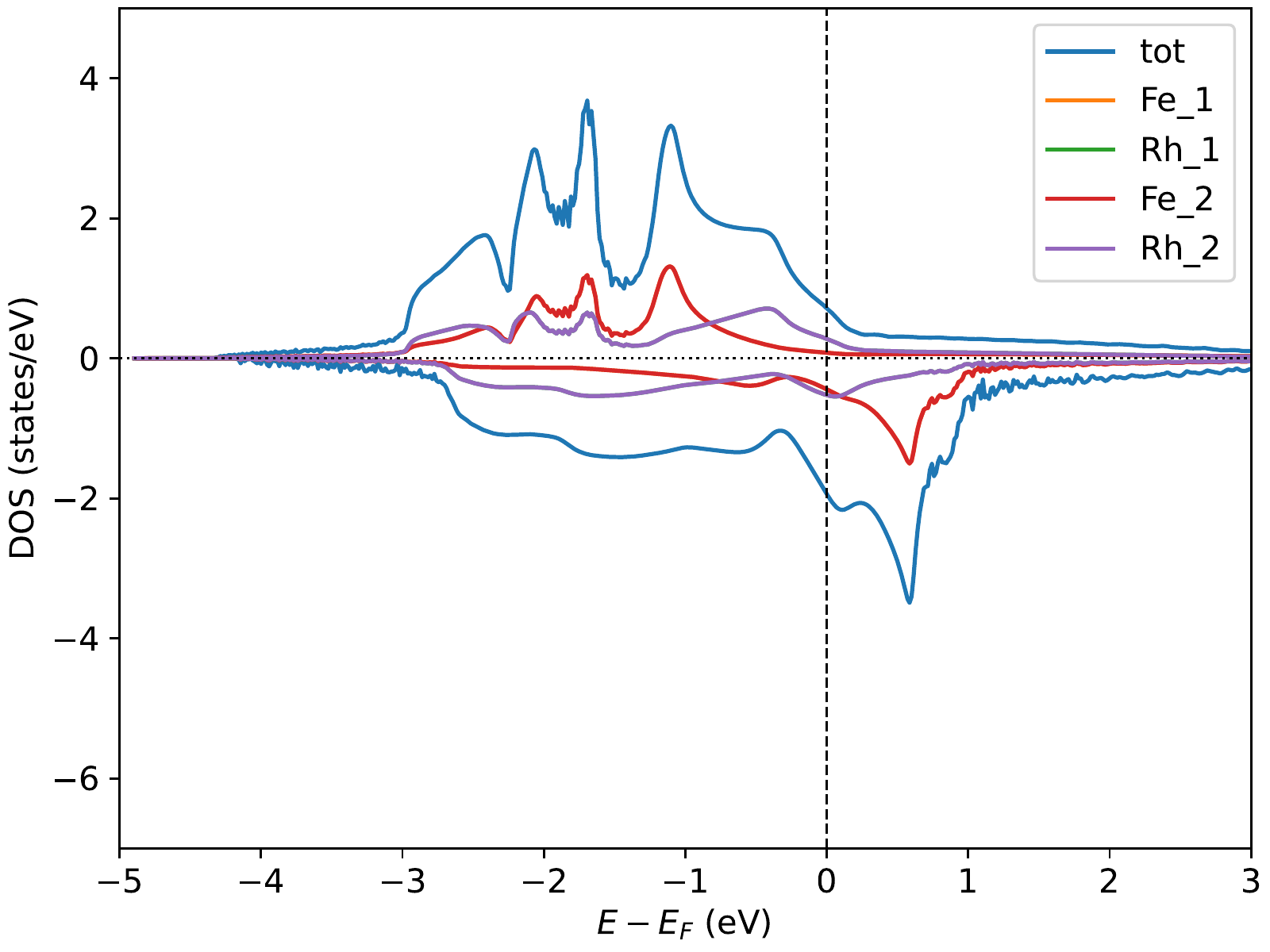}
			\caption{DOS for optimised lattice parameter($a$).}
			\label{fig:dos_d49_opt}
		\end{subfigure}
		\begin{subfigure}[b]{.3\columnwidth}
			\addtocounter{subfigure}{-1}
			\renewcommand\thesubfigure{\alph{subfigure}.3}
			\includegraphics[width=\columnwidth]{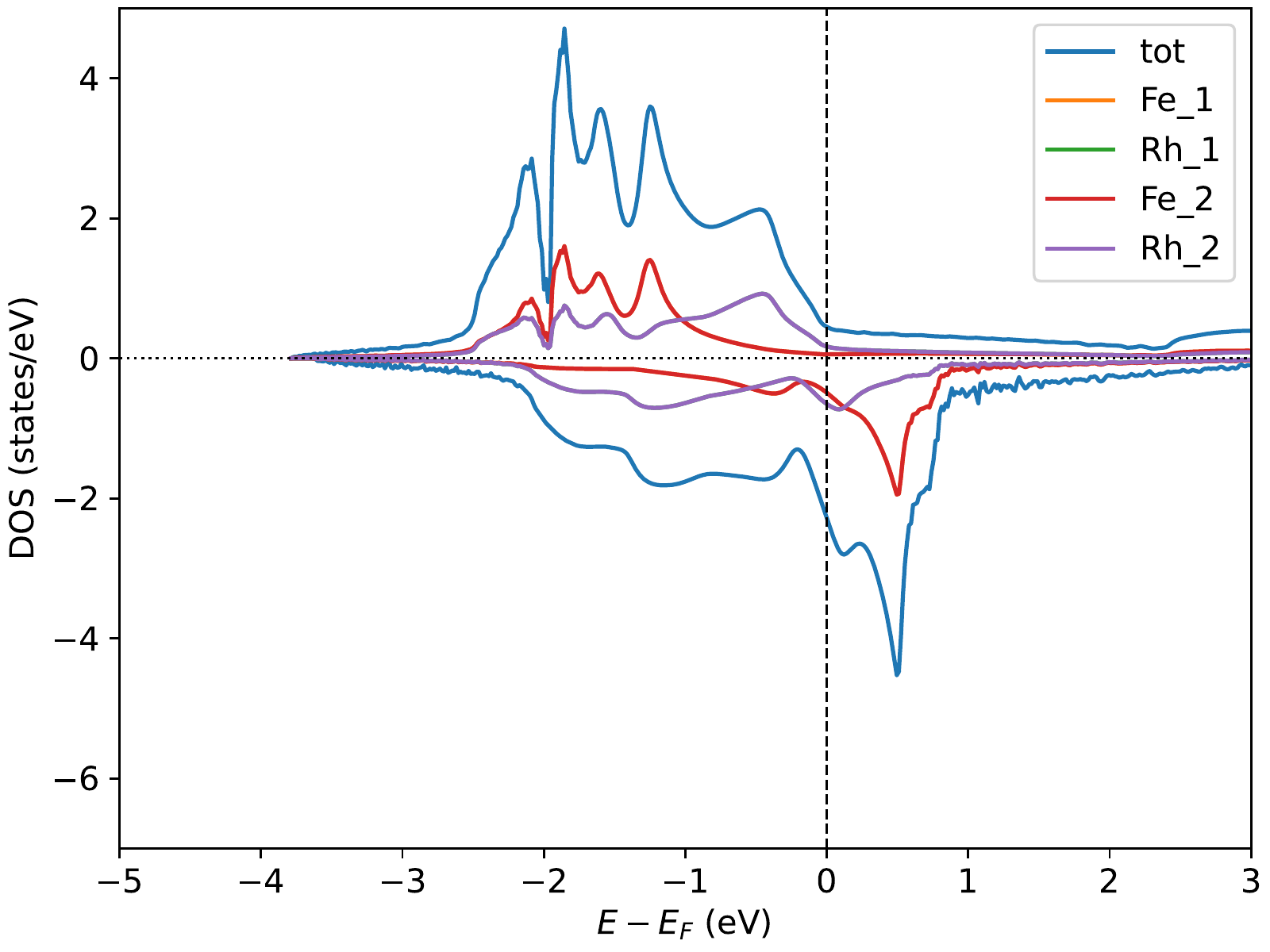}
			\caption{DOS for 6\% stretched lattice parameter($a$).}
			\label{fig:dos_d49_106}
		\end{subfigure}
		\addtocounter{subfigure}{-1}
		\caption{The DOS of disordered \fem.}
		\label{fig:dos_49d}
	\end{subfigure}
	\begin{subfigure}{\columnwidth}
		\begin{subfigure}[b]{.30\columnwidth}
			\renewcommand\thesubfigure{\alph{subfigure}.1}
			\includegraphics[width=\columnwidth]{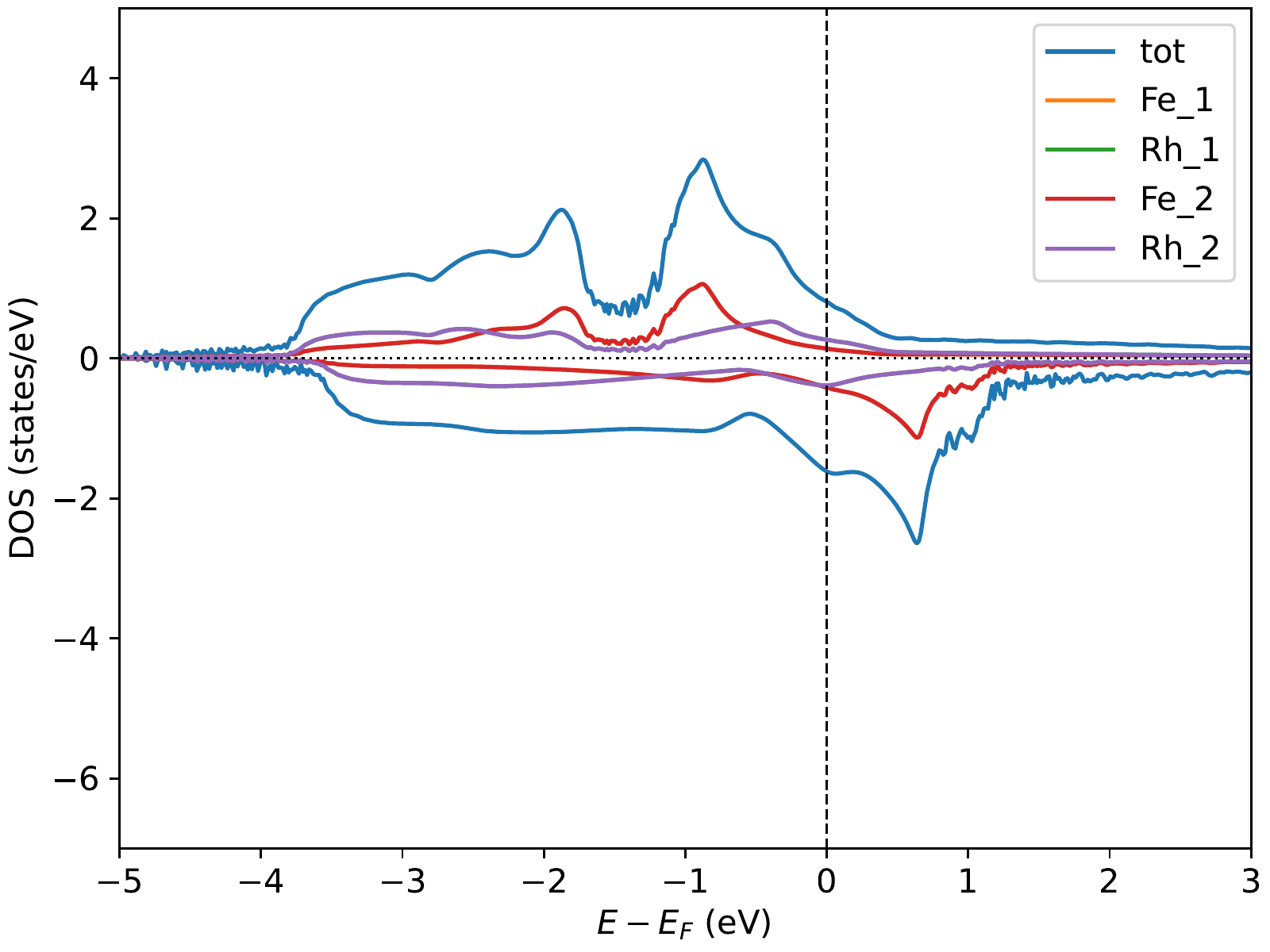}
			\caption{DOS for 6\% strained lattice parameter($a$).}
			\label{fig:dos_d50_94}
		\end{subfigure}
		\begin{subfigure}[b]{.30\columnwidth}
			\addtocounter{subfigure}{-1}
			\renewcommand\thesubfigure{\alph{subfigure}.2}
			\includegraphics[width=\columnwidth]{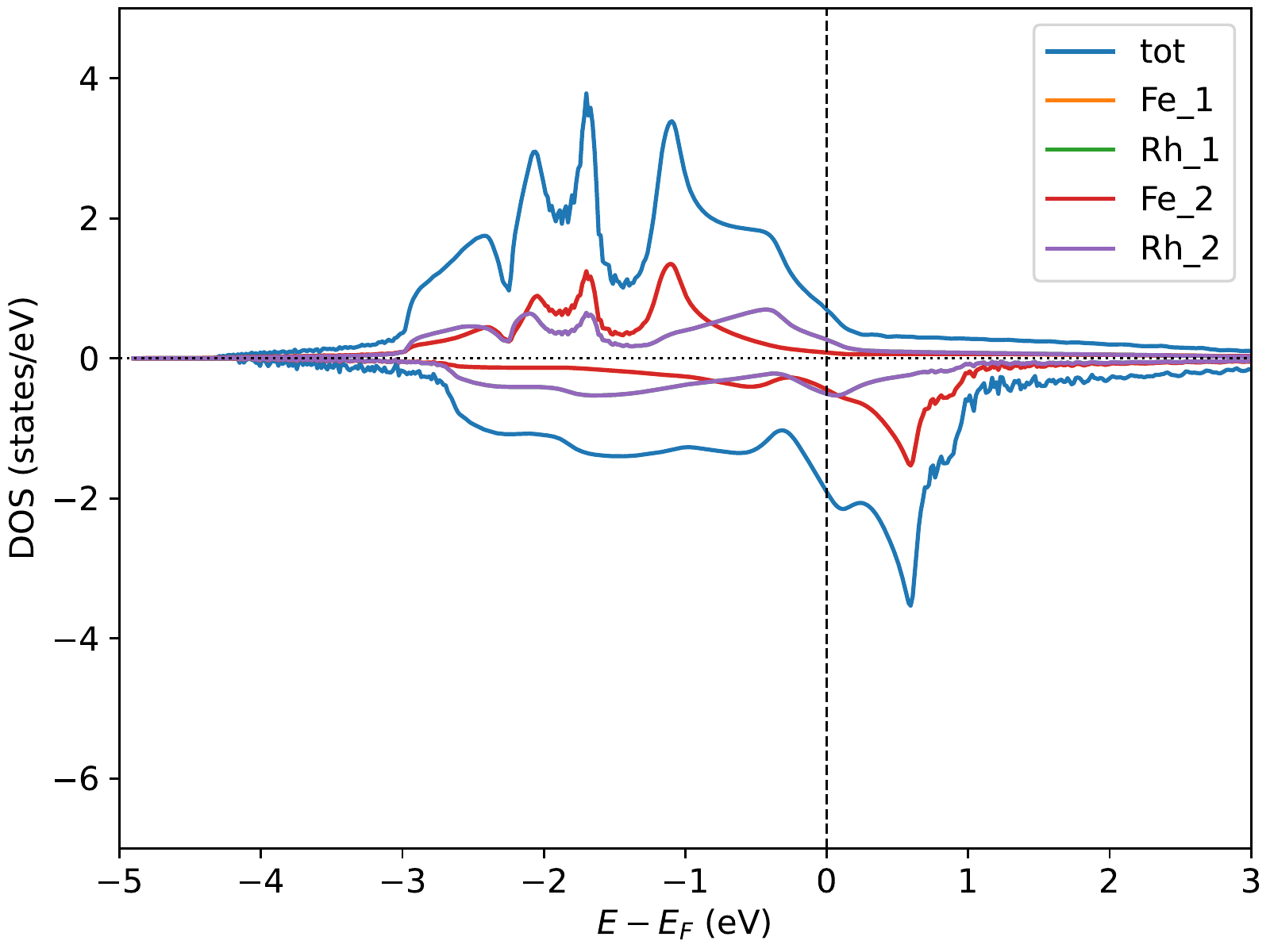}
			\caption{DOS for optimised lattice parameter($a$).}
			\label{fig:dos_d50_opt}
		\end{subfigure}
		\begin{subfigure}[b]{.3\columnwidth}
			\addtocounter{subfigure}{-1}
			\renewcommand\thesubfigure{\alph{subfigure}.3}
			\includegraphics[width=\columnwidth]{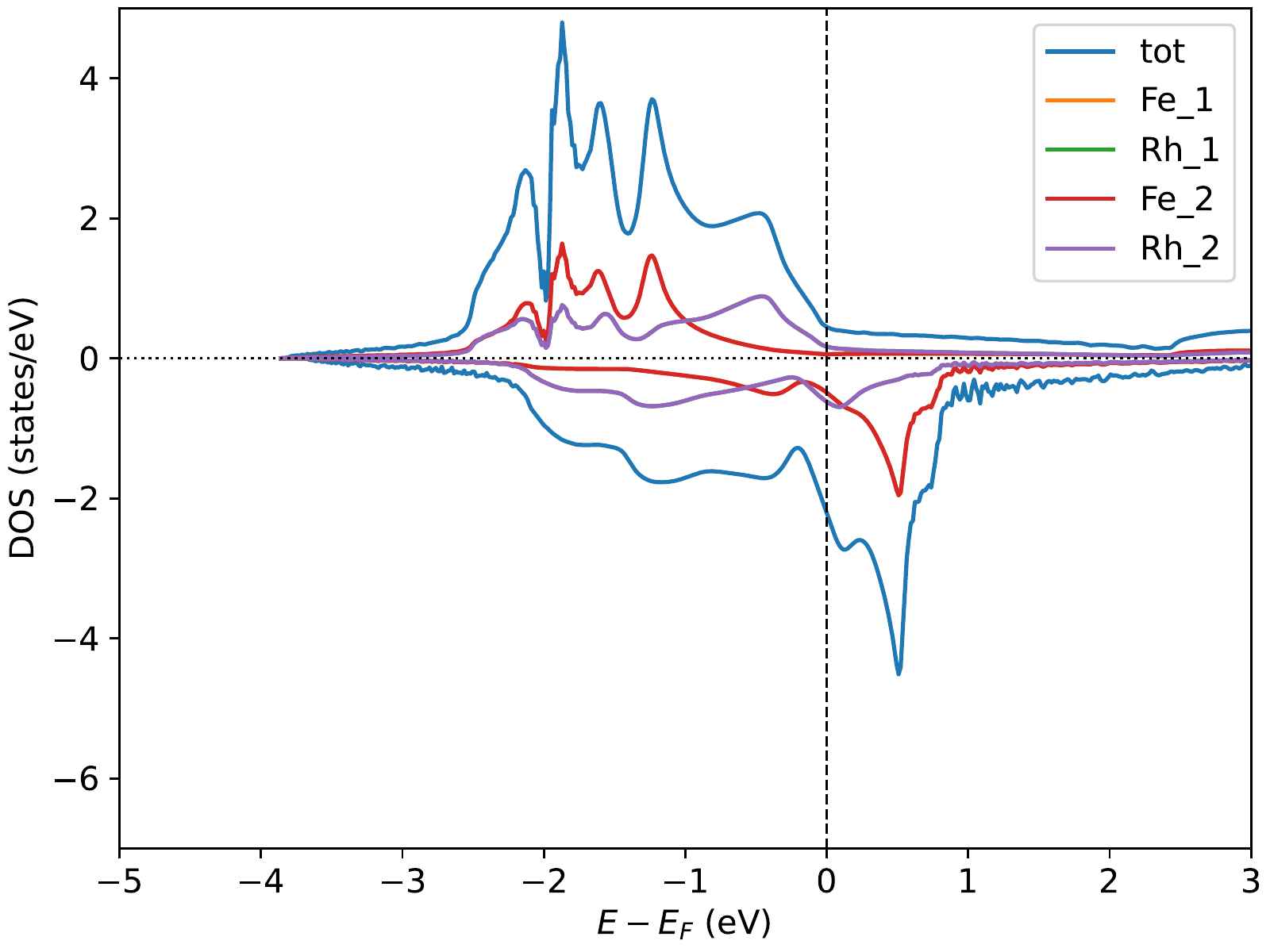}
			\caption{DOS for 6\% stretched lattice parameter($a$).}
			\label{fig:dos_d50_106}
		\end{subfigure}
		\addtocounter{subfigure}{-1}
		\caption{The DOS of disordered \feo.}
		\label{fig:dos_50d}
	\end{subfigure}
	\begin{subfigure}{\columnwidth}
		\begin{subfigure}[b]{.3\columnwidth}
			\renewcommand\thesubfigure{\alph{subfigure}.1}
			\includegraphics[width=\columnwidth]{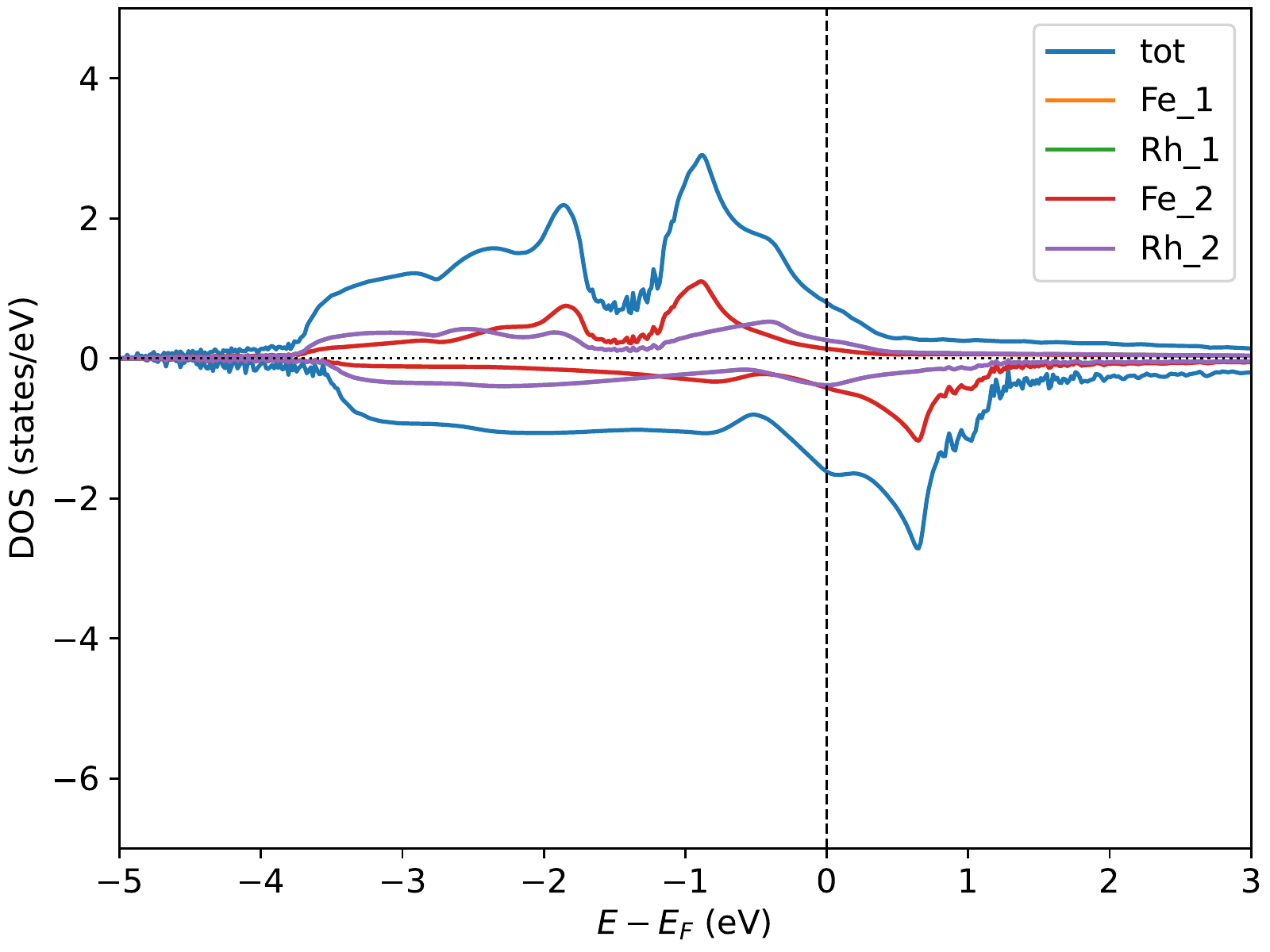}
			\caption{DOS for 6\% strained lattice parameter($a$).}
			\label{fig:dos_d51_94}
		\end{subfigure}
		\begin{subfigure}[b]{.3\columnwidth}
			\addtocounter{subfigure}{-1}
			\renewcommand\thesubfigure{\alph{subfigure}.2}
			\includegraphics[width=\columnwidth]{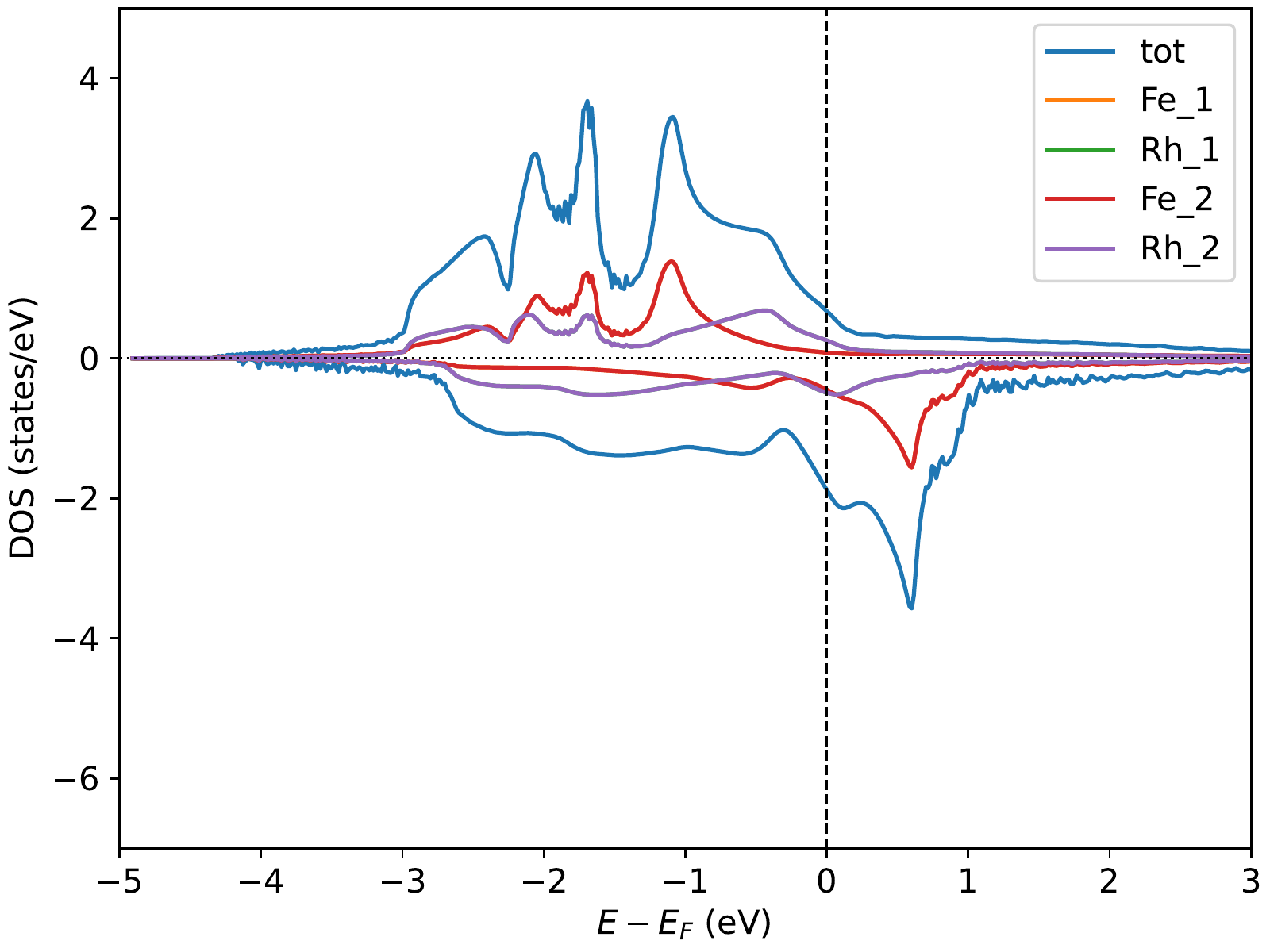}
			\caption{DOS for optimised lattice parameter($a$).}
			\label{fig:dos_d51_opt}
		\end{subfigure}
		\begin{subfigure}[b]{.3\columnwidth}
			\addtocounter{subfigure}{-1}
			\renewcommand\thesubfigure{\alph{subfigure}.3}
			\includegraphics[width=\columnwidth]{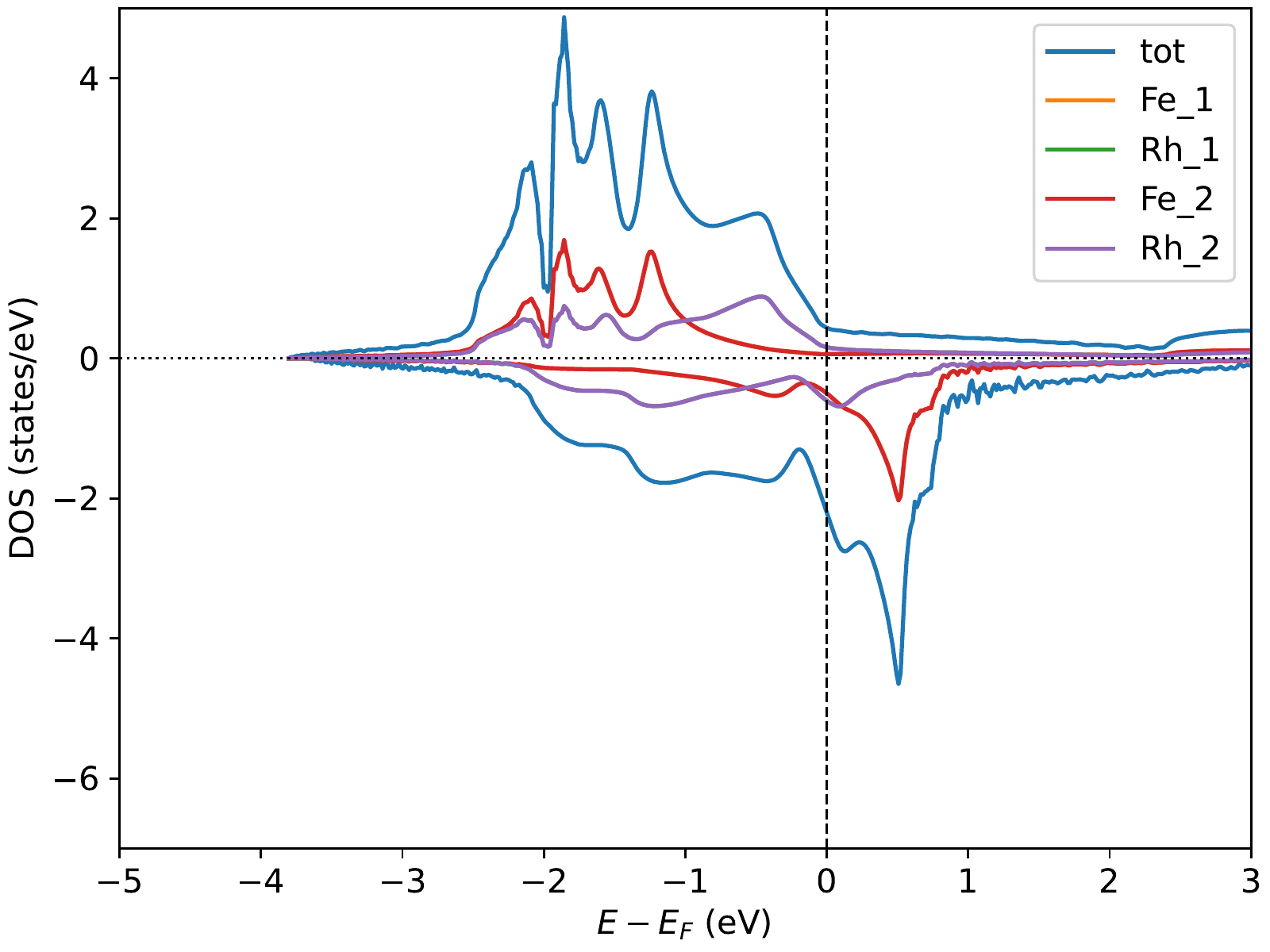}
			\caption{DOS for 6\% stretched lattice parameter($a$).}
			\label{fig:dos_d51_106}
		\end{subfigure}
		\addtocounter{subfigure}{-1}
		\caption{The DOS of disordered \fep.}
		\label{fig:dos_51d}
	\end{subfigure}
	\caption{Electronic structure of disordered \fex~systems for (\subref{fig:dos_49d})$x=$49,(\subref{fig:dos_50d})50, and
		(\subref{fig:dos_51d})51. While the electronic structures does not change significantly with $x$, it  does changes with
		lattice parameter $a$:
		(a) The width of the DOS squeezes with increasing $a$. (b) The spin-splitting in atom resolved DOS increases with $a$.}
	\label{fig:dos_d}
\end{figure}
In disordered systems \feo~shows bcc-A2 structure. In this case, Fe and Rh occupies both (0,0,0) and (.5,.5,.5) sites with
probability according to their concentrations. For brevity, we will denote Fe/Rh at (0,0,0) as Fe$_1$/Rh$_{1}$ and Fe/Rh at
(.5,.5,.5) as Fe$_2$/Rh$_{2}$.  It has to be noted that in this phase, there are 4 atoms in unit cell, unlike 2 atoms in section
(\ref{sec:ordered})
\subsubsection{Electronic structure}
\label{ssub:dis_electonic_structure}
The DOS for disordered phases are shown in the \figref{fig:dos_d}. The DOS is much smeared w.r.t. the ordered DOS plotted in
\figref{fig:ord_e}, which is a characteristic of disordered systems\cite{Kudrnovsk__2015,Banerjee_2010}.
Still, the main features of ordered state is still visible in disordered state as well. The valley near the -2eV in upspin
channel is the only distinct feature in strained state. The Rh DOS does not change significantly over the lattice parameter
ranges studied, the Fe atom starts showing more and more structures, bringing more features in total DOS. Beside the deep valley,
there is also a saddle neck just below the $E_F$ in major spin channel gets more and more visible. Likewise, the pseudo-gap
discussed in the section (\ref{ssub:ord_electronic_structure}) appears in both optimised and strained state.

\subsubsection{Magnetic Properties}
\label{ssub:magnetic_interactions}
\begin{figure}[ht!]
	\begin{subfigure}[t]{.3\columnwidth}
		\includegraphics[width=\columnwidth]{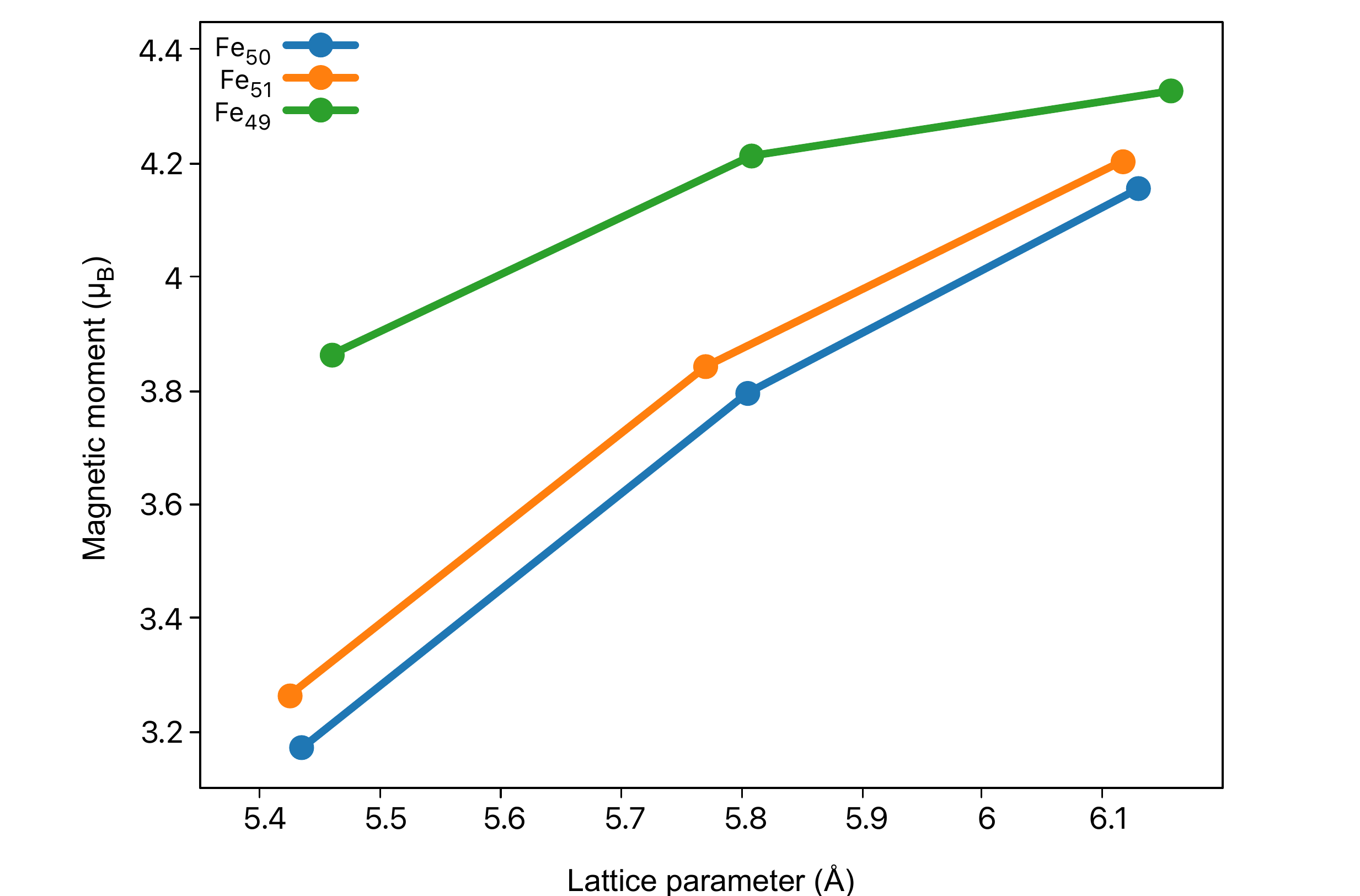}
		\caption{Magnetic moment of \fex.}
		\label{fig:mm_ferh}
	\end{subfigure}
	\begin{subfigure}[t]{.3\columnwidth}
		\includegraphics[width=\columnwidth]{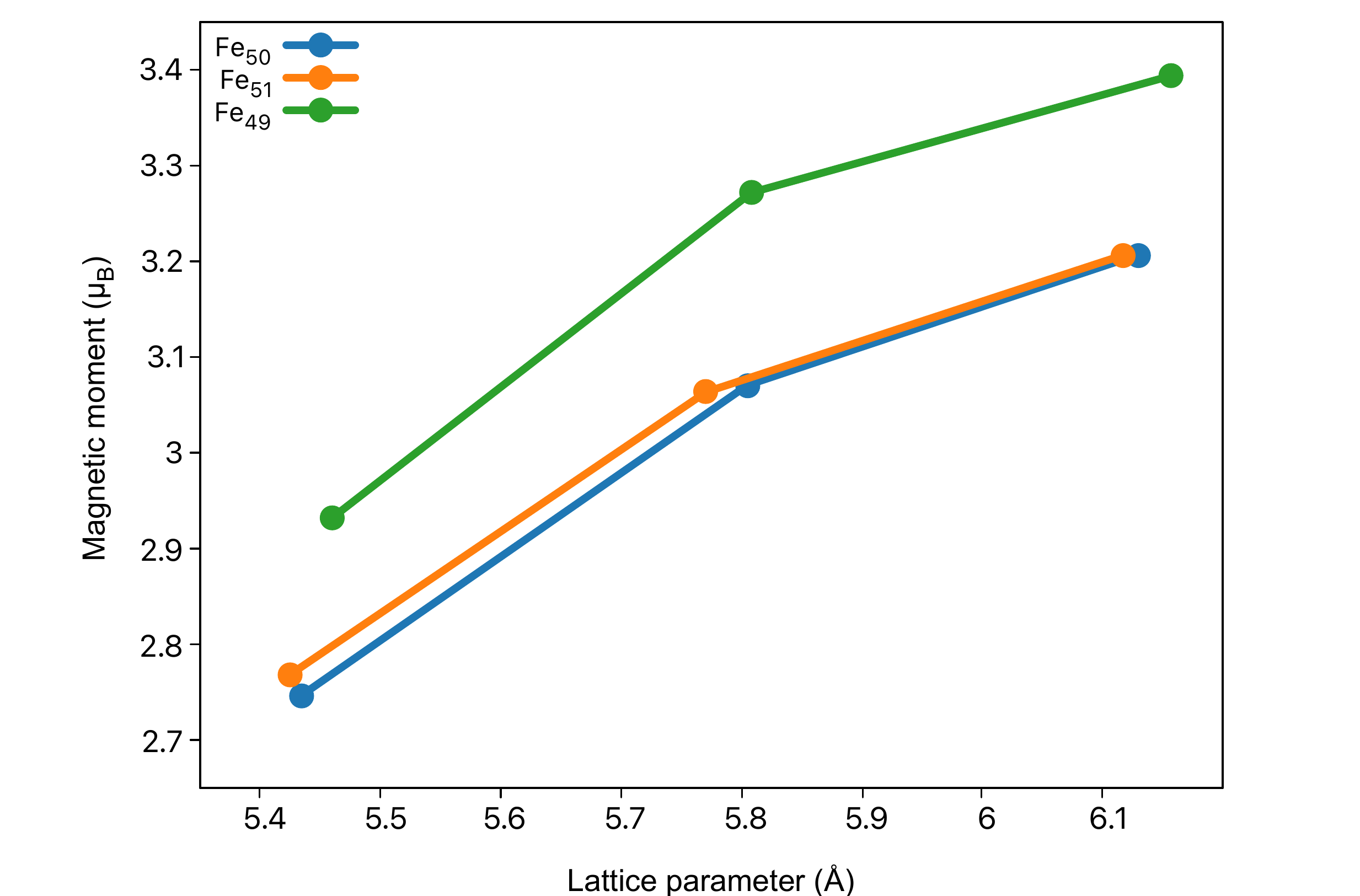}
		\caption{Magnetic moment of Fe atom in \fex.}
		\label{fig:mm_fe}
	\end{subfigure}
	\begin{subfigure}[t]{.3\columnwidth}
		\includegraphics[width=\columnwidth]{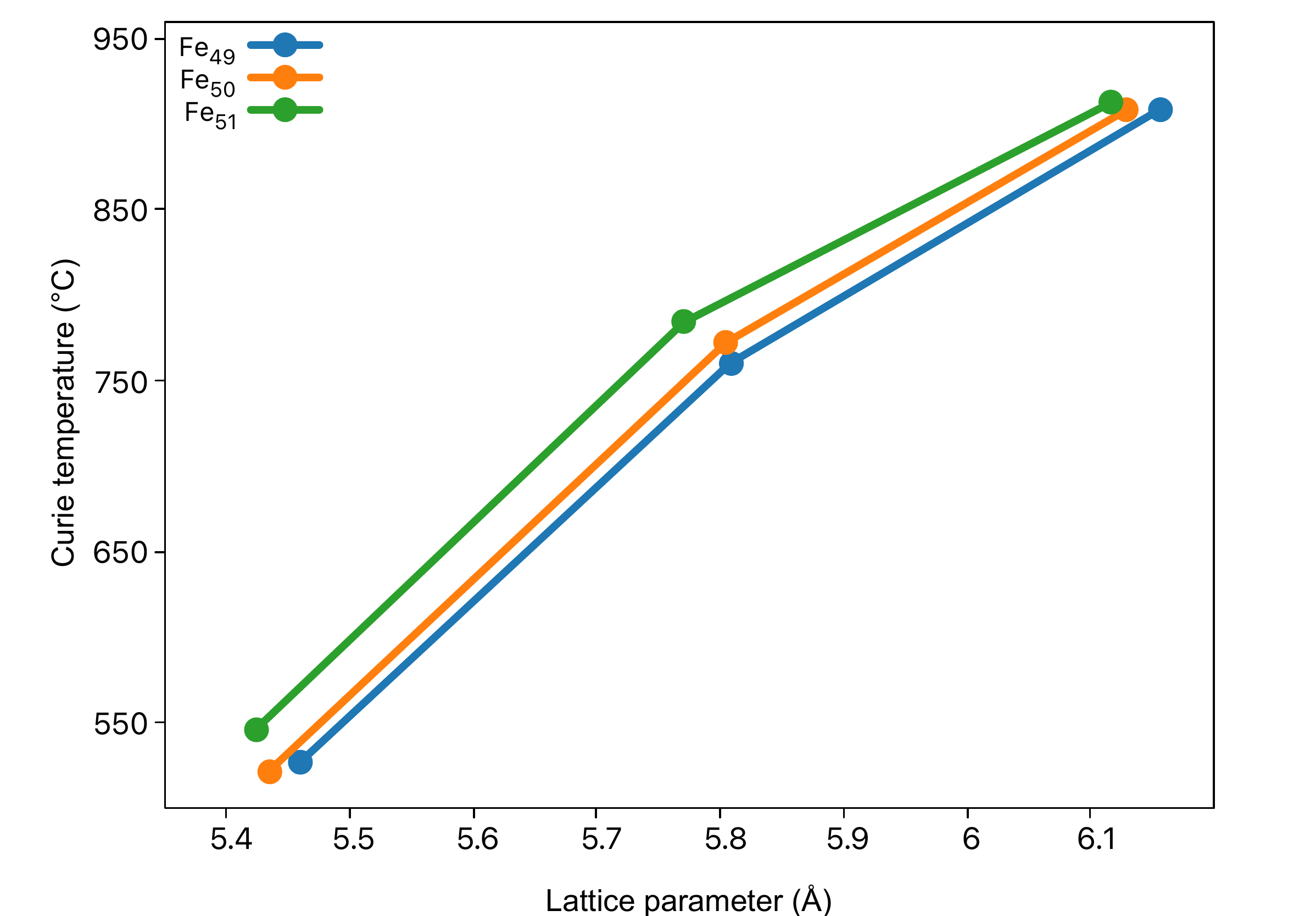}
		\caption{$T_C$ of \fex.}
		\label{fig:tc}
	\end{subfigure}
	\caption{Variation of Magnetic moment and Curie temperature of ordered \fex~with $x$ and $a$.}
	\label{fig:magmom}
\end{figure}

Similar to the electronic structures, magnetic interactions with Fe atom at the center does not change much with concentrations.
Unlike the ordered phase in section (\ref{ssub:ord_magnetic_interactions}), there is no visible spin glass phase in disordered
phase.
\begin{figure}[ht!]
	\centering
	\begin{subfigure}{\columnwidth}
		\begin{subfigure}[b]{.3\columnwidth}
			\renewcommand\thesubfigure{\alph{subfigure}.1}
			\includegraphics[width=\columnwidth]{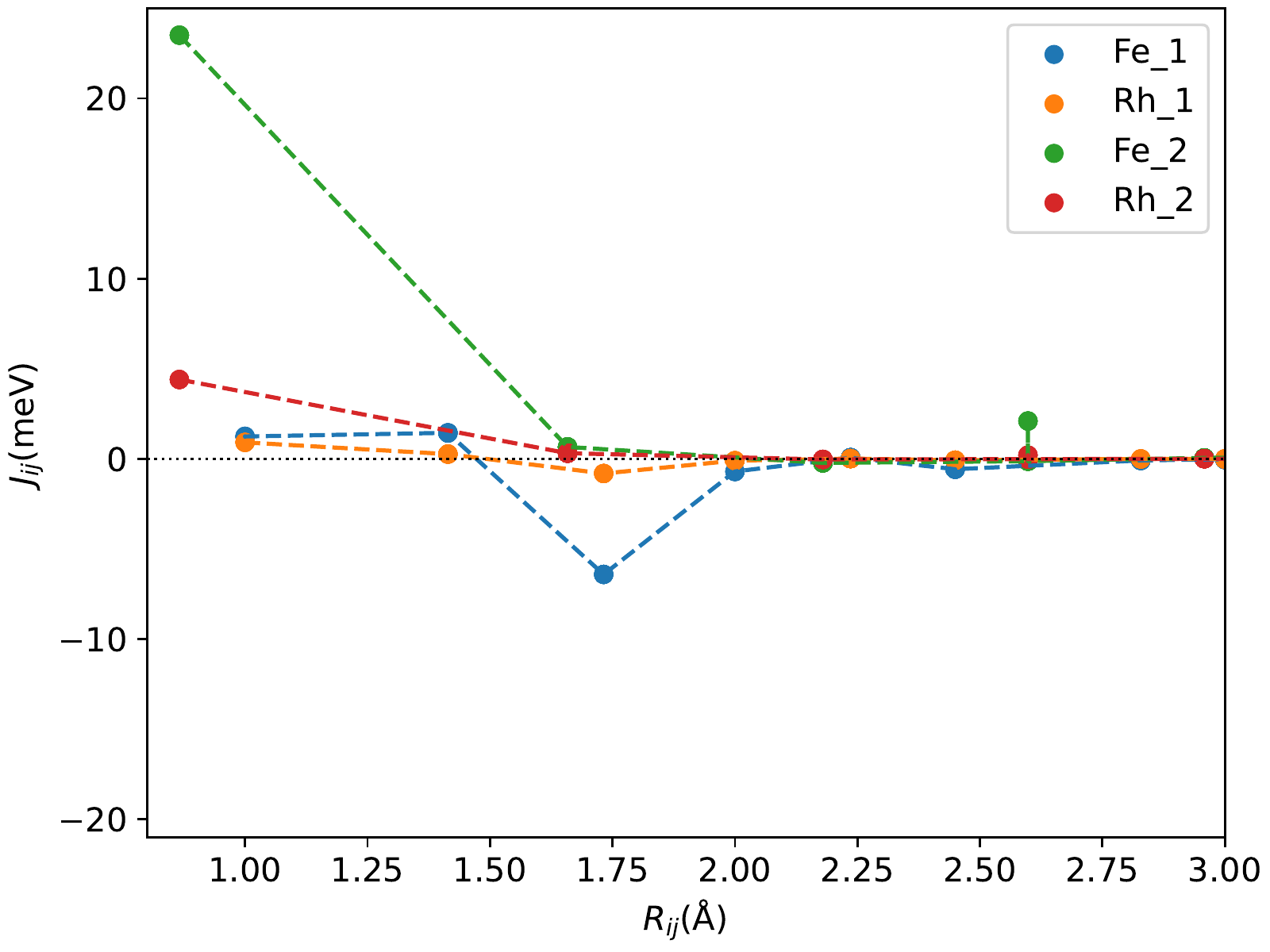}
			\caption{$\mathcal{J}_{ij}$ for 6\% strained lattice parameter($a$).}
			\label{fig:jij_d49_94}
		\end{subfigure}
		\begin{subfigure}[b]{.3\columnwidth}
			\addtocounter{subfigure}{-1}
			\renewcommand\thesubfigure{\alph{subfigure}.2}
			\includegraphics[width=\columnwidth]{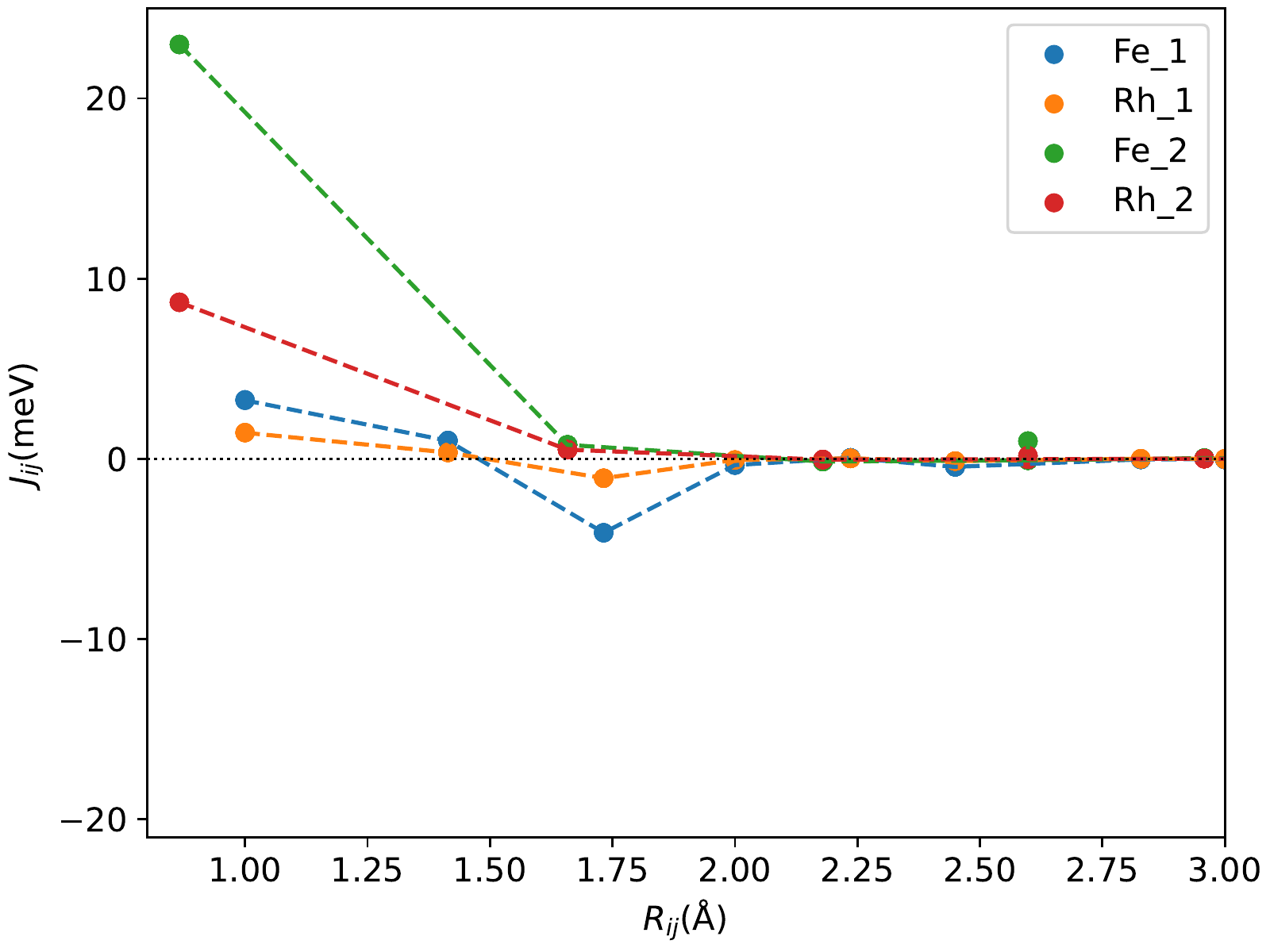}
			\caption{$\mathcal{J}_{ij}$ for optimised lattice parameter($a$).}
			\label{fig:jij_d49_opt}
		\end{subfigure}
		\begin{subfigure}[b]{.3\columnwidth}
			\addtocounter{subfigure}{-1}
			\renewcommand\thesubfigure{\alph{subfigure}.3}
			\includegraphics[width=\columnwidth]{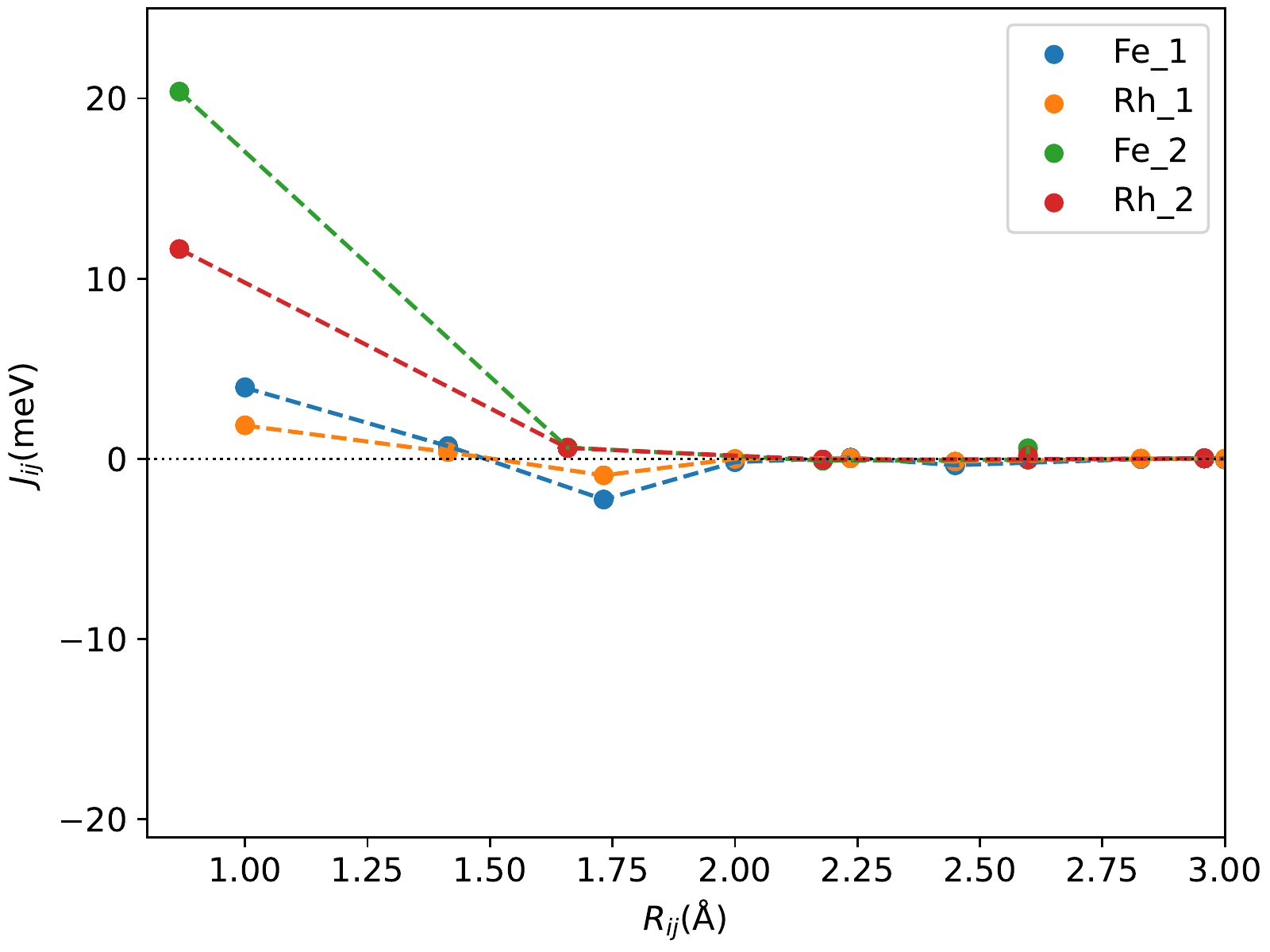}
			\caption{$\mathcal{J}_{ij}$ for 6\% stretched lattice parameter($a$).}
			\label{fig:jij_d49_106}
		\end{subfigure}
		\addtocounter{subfigure}{-1}
		\caption{The DOS and $\mathcal{J}_{ij}$ of disordered \fem.}
		\label{fig:jij_49d}
	\end{subfigure}

	\begin{subfigure}{\columnwidth}
		\begin{subfigure}[b]{.30\columnwidth}
			\renewcommand\thesubfigure{\alph{subfigure}.1}
			\includegraphics[width=\columnwidth]{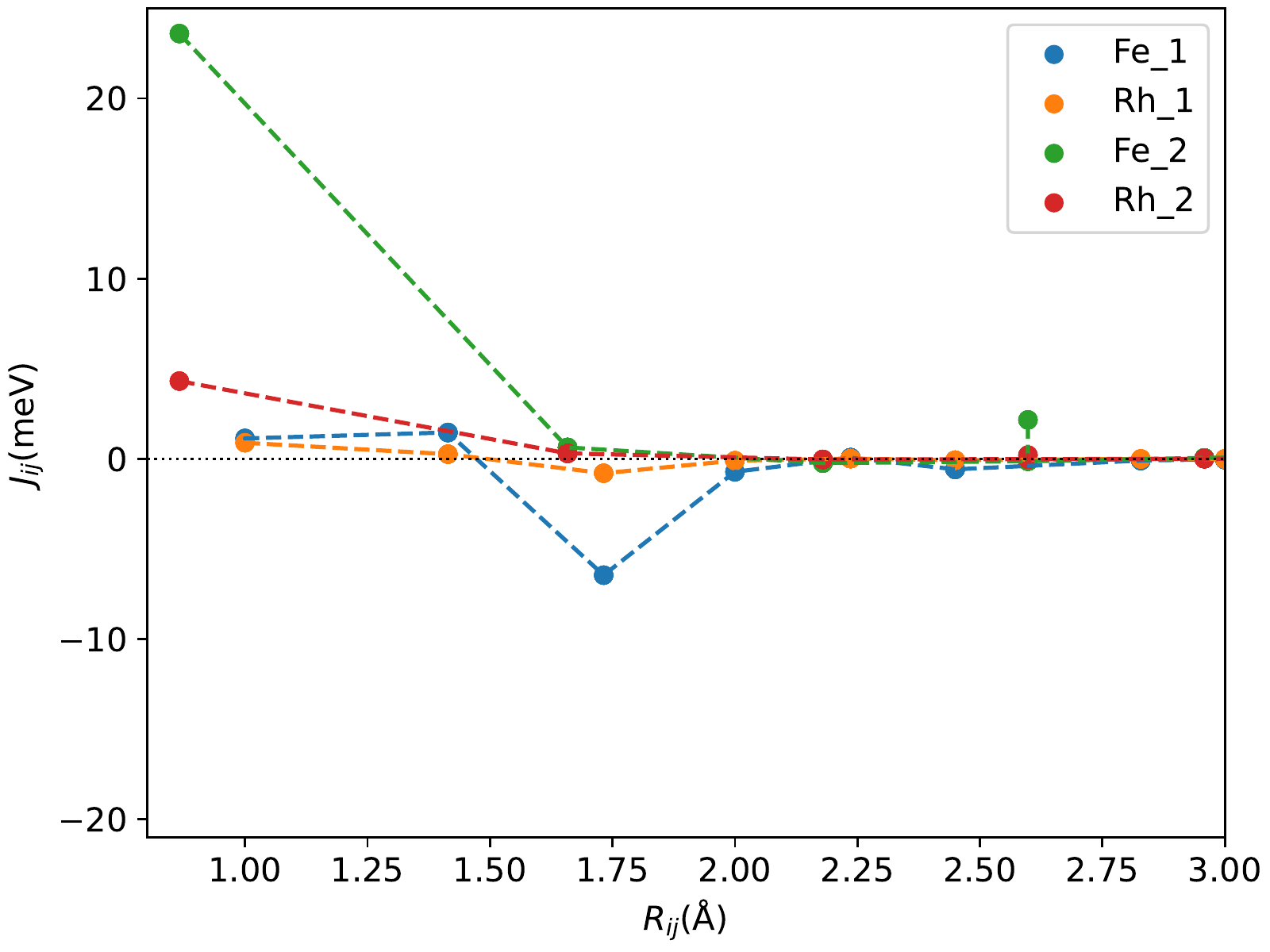}
			\caption{$\mathcal{J}_{ij}$ for 6\% strained lattice parameter($a$).}
			\label{fig:jij_d50_94}
		\end{subfigure}
		\begin{subfigure}[b]{.30\columnwidth}
			\addtocounter{subfigure}{-1}
			\renewcommand\thesubfigure{\alph{subfigure}.2}
			\includegraphics[width=\columnwidth]{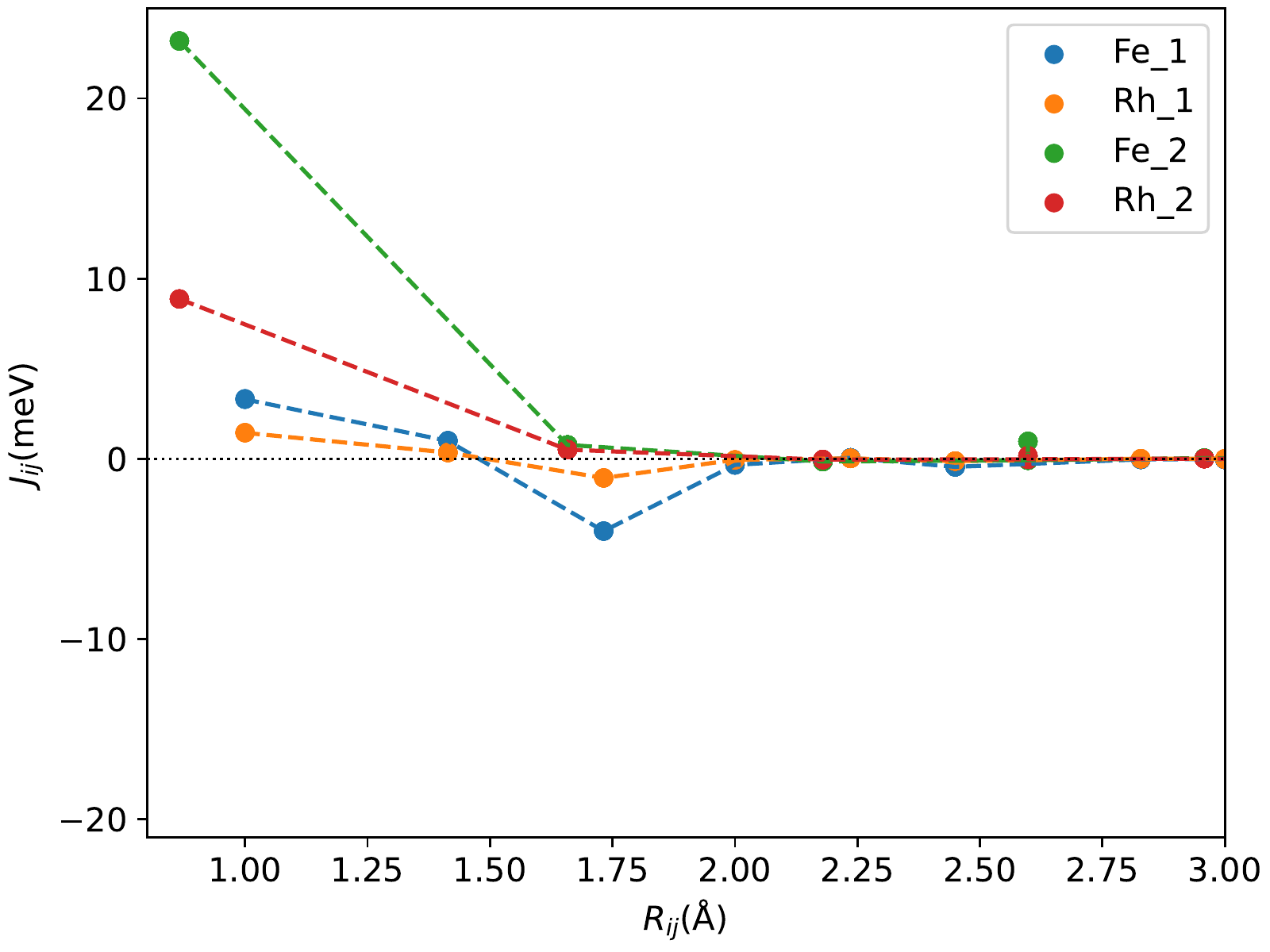}
			\caption{$\mathcal{J}_{ij}$ for optimised lattice pa  rameter($a$).}
			\label{fig:jij_d50_opt}
		\end{subfigure}
		\begin{subfigure}[b]{.3\columnwidth}
			\addtocounter{subfigure}{-1}
			\renewcommand\thesubfigure{\alph{subfigure}.3}
			\includegraphics[width=\columnwidth]{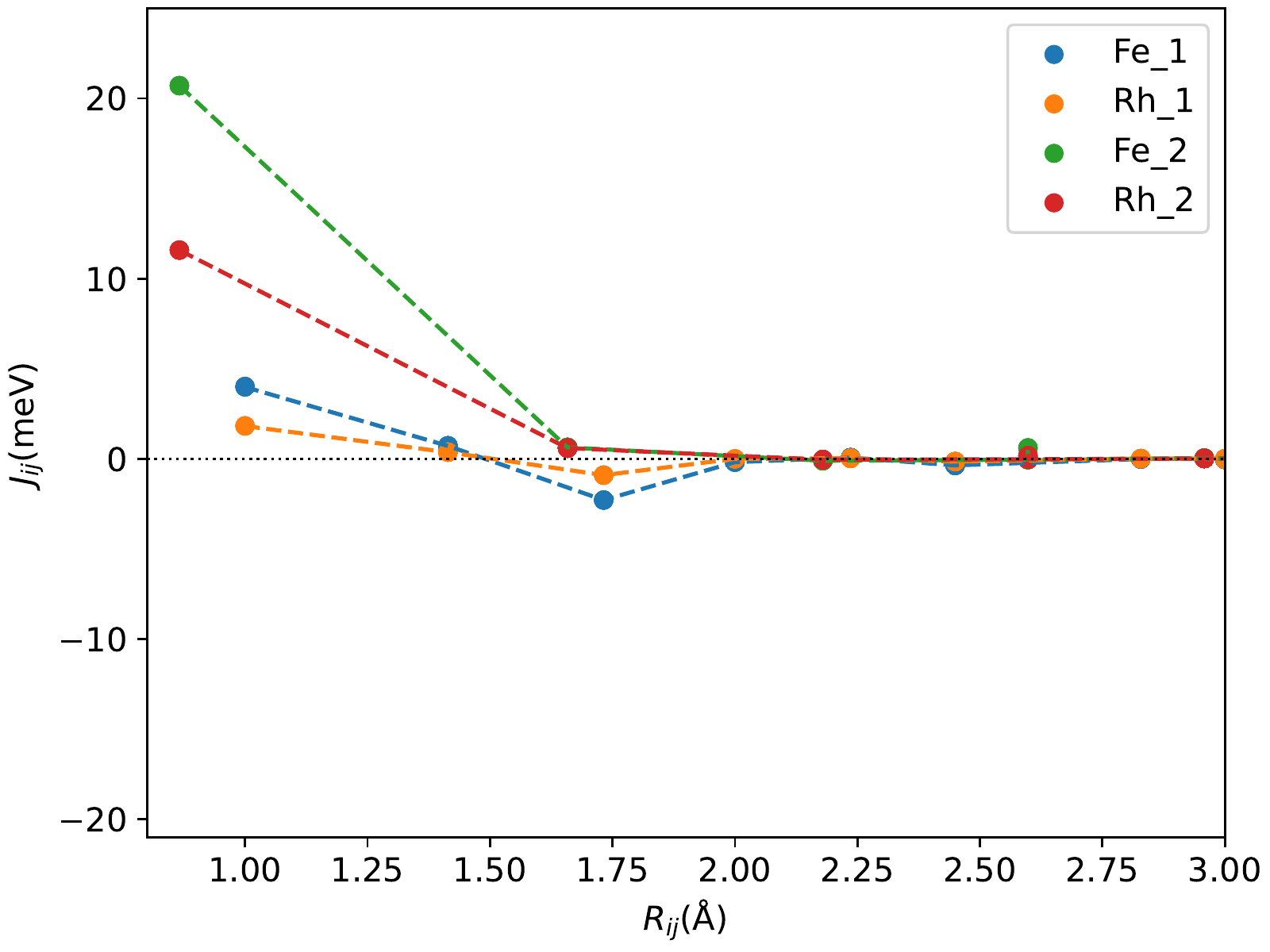}
			\caption{$\mathcal{J}_{ij}$ for 6\% stretched lattice parameter($a$).}
			\label{fig:jij_d50_106}
		\end{subfigure}
		\addtocounter{subfigure}{-1}
		\caption{The  $\mathcal{J}_{ij}$ of disordered \feo.}
		\label{fig:jij_50d}
	\end{subfigure}

	\begin{subfigure}{\columnwidth}
		\begin{subfigure}[b]{.3\columnwidth}
			\renewcommand\thesubfigure{\alph{subfigure}.1}
			\includegraphics[width=\columnwidth]{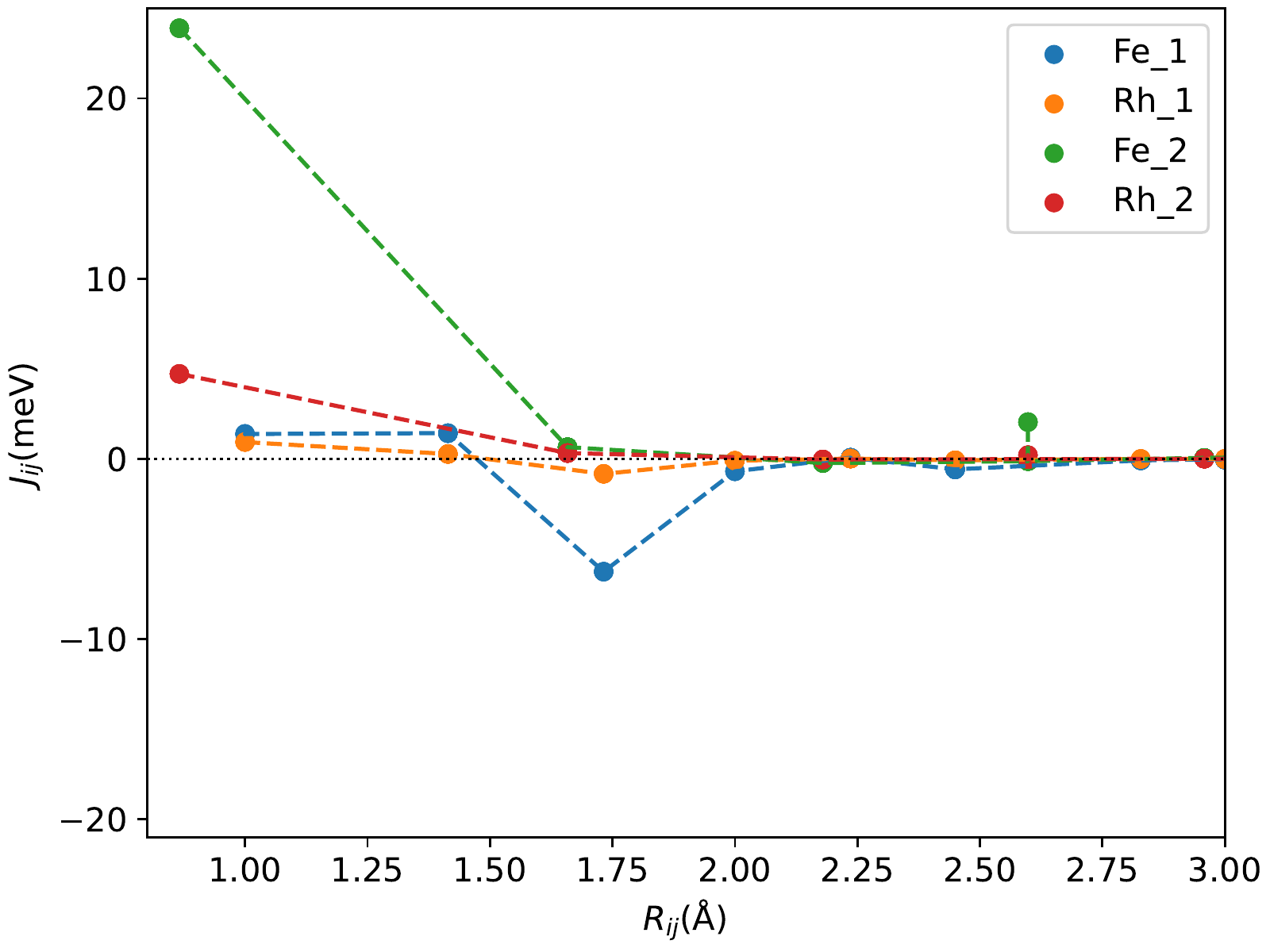}
			\caption{$\mathcal{J}_{ij}$ for 6\% strained lattice parameter($a$).}
			\label{fig:jij_d51_94}
		\end{subfigure}
		\begin{subfigure}[b]{.3\columnwidth}
			\addtocounter{subfigure}{-1}
			\renewcommand\thesubfigure{\alph{subfigure}.2}
			\includegraphics[width=\columnwidth]{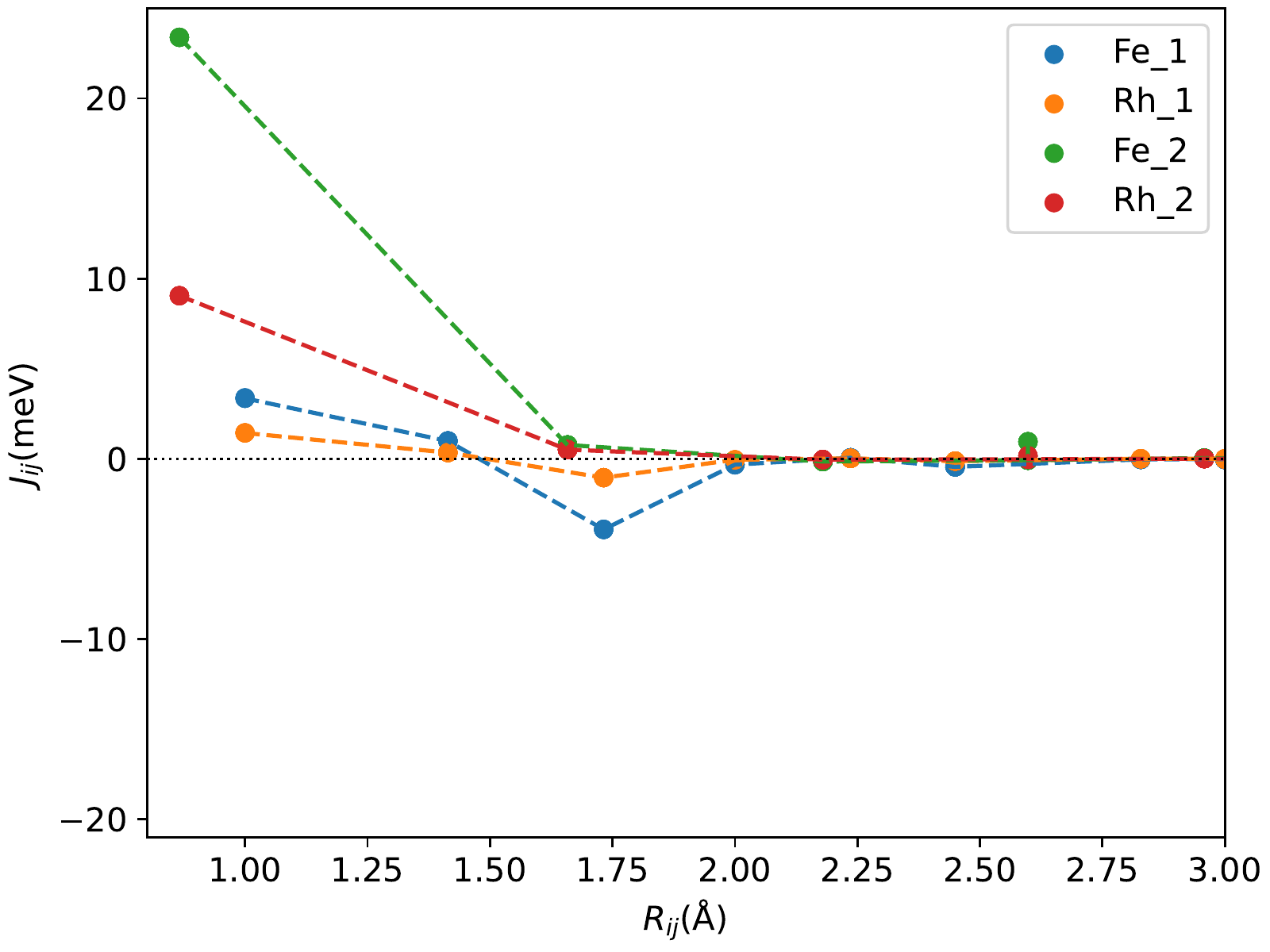}
			\caption{$\mathcal{J}_{ij}$ for optimised lattice parameter($a$).}
			\label{fig:jij_d51_opt}
		\end{subfigure}
		\begin{subfigure}[b]{.3\columnwidth}
			\addtocounter{subfigure}{-1}
			\renewcommand\thesubfigure{\alph{subfigure}.3}
			\includegraphics[width=\columnwidth]{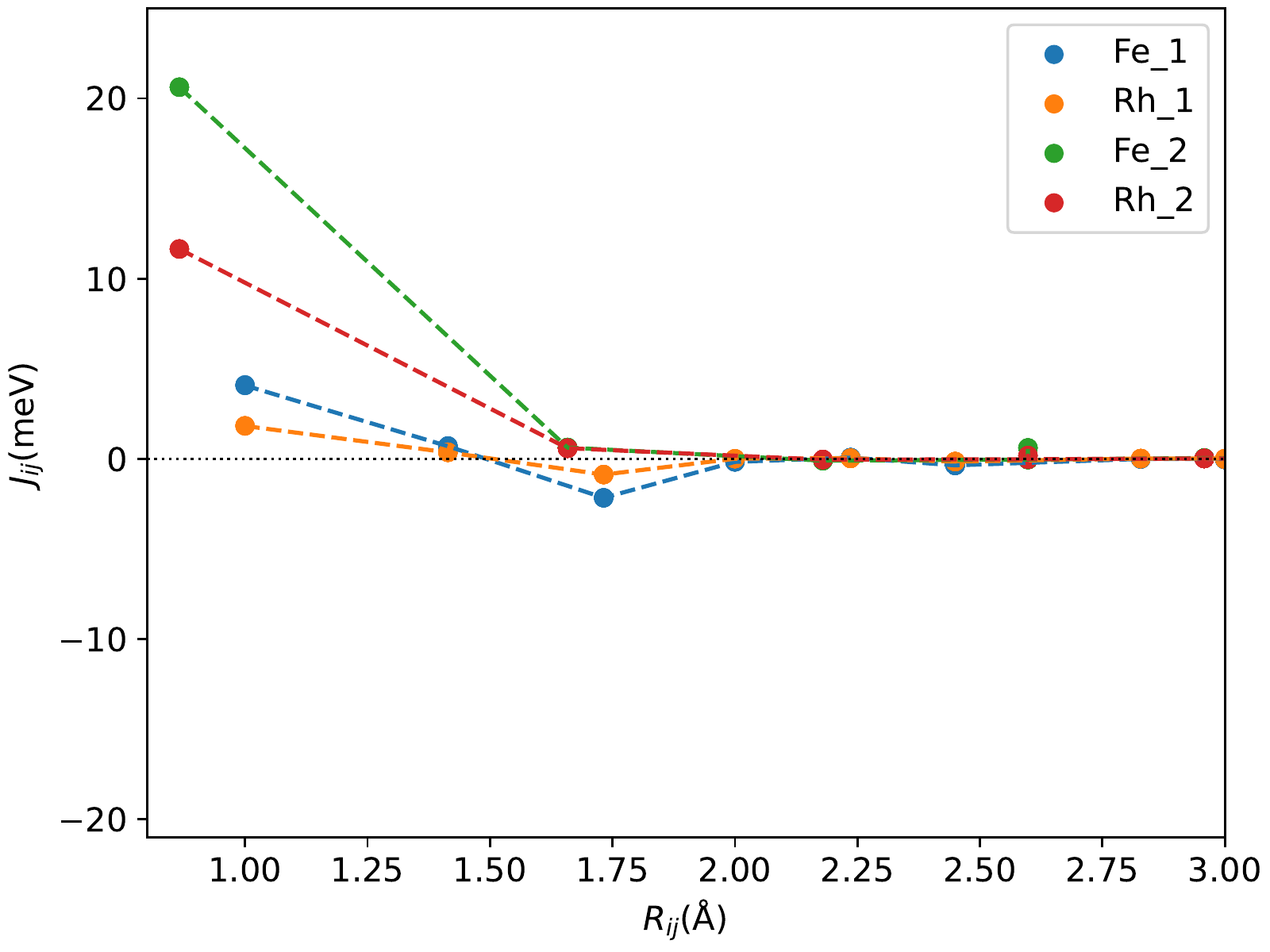}
			\caption{$\mathcal{J}_{ij}$ for 6\% stretched lattice parameter($a$).}
			\label{fig:jij_d51_106}
		\end{subfigure}
		\addtocounter{subfigure}{-1}
		\caption{The $\mathcal{J}_{ij}$ of disordered \fep.}
		\label{fig:jij_51d}
	\end{subfigure}
	\caption{\jij~of disordered \fex~systems for (\subref{fig:jij_49d})$x=$49, (\subref{fig:jij_50d})50 and
	(\subref{fig:jij_51d})51 with Fe$_1$ at the center.
	Similar to the electronic strictures, magnetic exchange interactions is not much changed with variations in $x$. The
	stress/strain, i.e. variation in $a$ changes the $\mathcal{J}_{ij}$ significantly. The $\mathcal{J}_0^{Fe}$ decreases
	slightly with increasing $a$, but $\mathcal{J}_0^{Rh}$ has increased to almost double with 12\% change in $a$.}
	\label{fig:jij_d}
\end{figure}
The variation in magnetic moments of \fex~and Fe atoms are shown in \figref{fig:mm_ferh} and \figref{fig:mm_fe}. The values are
tabulated in last three rows of the table (\ref{tab:magp}).  While the moment of Fe is very close to their atomic level (2.412
$\mu_B$ in 221 structure\cite{mp-568345}), the moment increases monotonously with the lattice parameter.  This trend shows
magnetic properties are more dependent on the distance between the atoms rather than the concentrations in this
off-stoichiometric systems. The magnetic moment of both Fe and Rh increase with the lattice parameter. This is due to the
increased spin-polarisation in both Fe and Rh, as seen in atom projected DOS in \figref{fig:dos_d}. This behaviour is completely
different from volume-preserved distortion, where $\mu_{Rh}$, $\mu_{Fe}$, $\mathcal{J}_0^{Rh}$ and $\mathcal{J}_0^{Fe}$ decreases
lattice parameter\cite{Witte_2016}. This indicates an establishment of long-range Fe-Fe FM interaction in this case, instead of a
competing short-range FM and long-range AFM interactions between Fe.

The Curie temperature calculated using MFT(\eqnref{eq:jij}) is tabulated in the table (\ref{tab:magp}) and
\figref{fig:tc}. Decreasing the Fe content also decreases the $T_C$. Our MFT result for B2-\feo~matches earlier results obtained
using same method. The lattice parameter effects $T_C$ profoundly, with around
70\% increase in $T_C$ for 12\% change in lattice parameter. The higher magnetic splitting in Rh has increased the
inter-sublattice interaction, leading to this high $\mathcal{J}_{ij}$, leading to the higher $T_C$.







\begin{table}[h!]
	\centering
	\begin{tabular}{cclcccc}
		\toprule
		\multirow{2}{*}{Fe}  & \multirow{2}{*}{Rh} & \multirow{2}{*}{} & \multicolumn{3}{c}{Magnetic moment ($\mu_B$)} &
		\multirow{2}{*}{$T_C(K)$}
		\\\cline{4-6}

		                     &                     &                   & Fe                                            & Rh     & FeRh   &     \\\midrule

		\multirow{3}{*}{50o} & 50                  & Stretched         & 2.9317                                        & 0.9319 & 3.8636 & 539 \\
		                     & 50                  & Optimised         & 3.2711                                        & 0.9417 & 4.2128 & 820 \\
		                     & 50                  & Strained          & 3.3929                                        & 0.9340 & 4.3269 & 832 \\\midrule
		\multirow{3}{*}{49}  & 51                  & Stretched         & 2.7786                                        & 0.4301 & 3.1616 & 522 \\
		                     & 51                  & Optimised         & 3.0774                                        & 0.7173 & 3.7475 & 772 \\
		                     & 51                  & Strained          & 3.2200                                        & 0.9620 & 4.1369 & 907 \\\midrule
		\multirow{3}{*}{50d} & 50                  & Stretched         & 2.7472                                        & 0.4239 & 3.1711 & 527 \\
		                     & 50                  & Optimised         & 3.0703                                        & 0.7253 & 3.7956 & 760 \\
		                     & 50                  & Strained          & 3.2062                                        & 0.9480 & 4.1542 & 908 \\\midrule
		\multirow{3}{*}{51}  & 49                  & Stretched         & 2.7679                                        & 0.4494 & 3.2637 & 546 \\
		                     & 49                  & Optimised         & 3.0631                                        & 0.7331 & 3.8427 & 784 \\
		                     & 49                  & Strained          & 3.2058                                        & 0.9531 & 4.2042 & 912 \\

		\bottomrule
	\end{tabular}
	\caption{Magnetic properties of \fex. The 50o and 50d in first and third row  means the results are for ordered and disordered
		\feo.}
	\label{tab:magp}
\end{table}

\section{Conclusion} 
In this communication, we have presented, electronic and magnetic properties of \fex~as obtained from density functional theory
based calculations. The main outcome of the calculations can be summarised as
\begin{enumerate*}[label=(\roman*)]
	\item The electronic structures of \fex~matches well with previous calculations.
	\item The magnetic moment of both Fe and Rh is higher than the free atoms. The magnetic moment of Rh in disordered phase is due
	to the hybridisation of nearest neighbour Fe.
	\item Both electronic and magnetic properties has stronger dependency on the distance between nearest neighbours or the lattice
	parameters than the small variations in composition.
\end{enumerate*}
Particularly, we have shown the effect of chemical disorder and pressure on electronic and magnetic properties of \fex both
separately and together. This will help further understanding and designing Fe-based caloric materials.

\section{Acknowledgment}
\label{sec:acknowledgment}
We acknowledge the High Performance Computing Center (HPCC), SRM IST for providing the computational facility to carry out this
research work successfully.

\bibliographystyle{unsrt}
\bibliography{ferh}

\end{document}